\documentclass{aa}  

\usepackage{graphicx}
\usepackage{txfonts}

\usepackage{lscape}
\usepackage{longtable}
\usepackage[colorlinks=true,     linkcolor=blue, citecolor=blue, filecolor=blue, urlcolor=blue]{hyperref}
\usepackage{xparse}
\usepackage[normalem]{ulem}
\usepackage{multirow}

\usepackage{listings}

\begin{document} 

   \title{Rendezvous in CAVITY: Kinematics and gas properties of an isolated dwarf-dwarf merging pair in a cosmic void region}
   \titlerunning{An isolated merging pair of dwarf galaxies in void}

   \author{Bahar Bidaran\inst{\ref{ugr1}}
   \and Simon de Daniloff\inst{\ref{ugr1},\ref{IRAM}}
   \and Isabel Pérez\inst{\ref{ugr1},\ref{ugr2}}
    \and Almudena Zurita\inst{\ref{ugr1},\ref{ugr2}}  
    \and Javier Román\inst{\ref{Complutense}}
    \and María Argudo-Fernández\inst{\ref{ugr1},\ref{ugr2}}
    \and Daniel Espada\inst{\ref{ugr1},\ref{ugr2}}
    \and Tomás Ruiz-Lara\inst{\ref{ugr1},\ref{ugr2}}
   \and Laura Sánchez-Menguiano\inst{\ref{ugr1},\ref{ugr2}}
    \and Rubén García-Benito \inst{\ref{iaa}}
    \and Reynier F. Peletier\inst{\ref{kapteyn}}
    \and Anna Ferr\'e-Mateu\inst{\ref{iac},\ref{ull}}
     \and Salvador Duarte Puertas\inst{\ref{ugr1},\ref{ugr2},\ref{laval}}
    \and Simon Verley\inst{\ref{ugr1},\ref{ugr2}}
     \and Jes\'us Falc\'on-Barroso\inst{\ref{iac},\ref{ull}}
    \and Estrella Florido\inst{\ref{ugr1},\ref{ugr2}}
     \and Gloria Torres-Ríos\inst{\ref{ugr1}}
    \and Ute Lisenfeld\inst{\ref{ugr1},\ref{ugr2}}
    \and Mónica Relaño\inst{\ref{ugr1},\ref{ugr2}}
    \and Andoni Jiménez\inst{\ref{ugr1}}
    }

   \institute{
   Dpto. de F\'{\i}sica Te\'orica y del Cosmos, Facultad de Ciencias (Edificio Mecenas), University of Granada, E-18071, Granada, Spain\label{ugr1}
    \and Institut de Radioastonomie Millim\'etrique (IRAM), Av. Divina Pastora 7, N\'ucleo Central 18012, Granada, Spain \label{IRAM}
    \and Instituto Carlos I de F\'\i sica Te\'orica y Computacional, Universidad de Granada, 18071, Granada, Spain\label{ugr2}
    \and Departamento de Física de la Tierra y Astrofísica, Universidad Complutense de Madrid, 28040 Madrid, Spain \label{Complutense}
    \and Instituto de Astrof\'isica de Andaluc\'ia - CSIC, Glorieta de la Astronomía s/n, 18008 Granada, Spain\label{iaa}          
    \and Kapteyn Astronomical Institute, University of Groningen, PO Box 800, 9700 AV Groningen, The Netherlands\label{kapteyn}
    \and Instituto de Astrof\'isica de Canarias, c/V\'ia L\'actea s/n, E-38205, La Laguna, Tenerife, Spain\label{iac}
    \and Departamento de Astrof\'isica, Universidad de La Laguna, E-38206, La Laguna, Tenerife, Spain\label{ull}
    \and D\'epartement de Physique, de G\'enie Physique et d’Optique, Universit\'e Laval, and Centre de Recherche en Astrophysique du Qu\'ebec (CRAQ), Québec, QC, G1V 0A6, Canada\label{laval}
    }
    
   \date{Received Month Day, Year; accepted Month Day, Year}

  \abstract
   {Galaxy mergers are pivotal events in the evolutionary history of galaxies, with their impact believed to be particularly significant in dwarf galaxies due to their low masses. However, these events remain largely underexplored, especially in pristine environments such as voids. }
   {In this work, we report the serendipitous identification of an isolated merging dwarf system with a total stellar mass of M$_{\rm \star}$~$\sim$~10$^{9.7}$~M$_{\rm \odot}$, located in the centre of a cosmic void. {This system is one of the rare examples, and possibly the first, of merging dwarf galaxy pairs studied within the central region of a cosmic void}. This system is remarkable due to its orientation relative to the line of sight and its unique local and large-scale environment.}
   {Using CAVITY PPAK-IFU data combined with deep optical broadband imaging from the Isaac Newton Telescope, we analysed the kinematics and ionised gas properties of each dwarf galaxy in the system by employing a full spectral fitting technique.}
   {The orientation of this merging pair relative to the line of sight allowed us to determine the dynamical mass of each component, which we found to have similar dynamical masses within galactocentric distances of up to {2.9 kpc}. These galaxies were likely star-forming dwarfs with rotating discs prior to the merger. {While the gas-phase metallicity of both components is consistent with that of star-forming dwarf galaxies,} the star formation rates observed in both components exceed those typically reported for equally massive star-forming dwarf galaxies. This indicates that the merger has presumably contributed to enhancing star formation. {Our analysis shows no signs of AGN activity in this merging pair. }Furthermore, we found no significant difference in the optical \textit{g-r} colour of this merging pair compared to other merging dwarf pairs across different environments.}
   {While most merging events occur in group-like environments with a high galaxy density and the tidal influence of a host halo, and isolated mergers typically involve galaxies with significant mass differences, the identified merging pair does not follow these patterns. We speculate that the global dynamics of the void or past three-body encounters involving components of this pair and a nearby dwarf galaxy might have triggered this merging event.
}


   \maketitle

\section{Introduction}\label{Introduction}

Merger events play a crucial role in the mass growth and evolution of galaxies, as they significantly impact their physical properties, including their star formation rate (SFR), morphology, and black hole growth \citep[e.g.][]{1972Toomre, 2010Hopkins}. The hierarchical $\Lambda$ cold dark matter ($\Lambda$CDM) paradigm \citep{2000Cole} predicts that merger events are frequent throughout the evolutionary history of galaxies, occurring commonly among low-mass ones with log(M$_{\star}$/M$_{\odot}$) $<$ 9.5 \citep[e.g.][]{2010Klimentowski,2014Deason}. Recent observations increasingly support the prevalence of such events among dwarf galaxies \cite[e.g.][]{2012ApJ...748L..24M,2012Rich,2014Amorisco,2014ApJ...795L..35C}. Also, N-body simulations, such as Millennium-II, predict at least three major mergers\footnote{Major mergers are mergers between galaxies of comparable stellar masses.} for dwarf galaxies during their lifetime \citep{2010Fakhouri}. \cite{2014Deason} shows that more than 10$\%$ of dwarf galaxies, such as satellites of the Milky Way (MW) or M31-like host halos, underwent at least one major merger event since redshift $z=1$, and prior to infall to their present-day host halo or shortly ($\sim$ 1-3 Gyr) after that \citep[see also][]{2015Wetzel}. Beyond the virial radius of the MW or M31-like host halos, and in more isolated environments, the anticipated number of recent merger events doubles. \cite{2021Martin} shows that isolated dwarf galaxies and those in galaxy groups had experienced at least one minor and one major merger during their lifetime between redshift 0.5$<$ z $<$5.0.

The impact of merger events is expected to be particularly pronounced in dwarf galaxies due to their low mass. 
Minor and major mergers, along with internal processes and tidal interactions, are believed to be the primary drivers of the permanent morphological transformations in dwarfs \citep[e.g.][]{1977Toomre,2006Knebe,2009Dekel,2014Naab,2018Martin}. Furthermore, mergers are believed to be the main trigger of intense star formation in pairs \citep[e.g.][]{2002Tissera, 2013Ellison}, some of which result in the formation of starburst dwarfs, such as blue compact dwarf galaxies \citep[BCDs;][]{2020Zhang}. Similarly, {\cite{2021Ruiz-Lara} interprets {the reignition of star formation after a quiescent period} and the presence of metal-poor stars in Leo I as evidence of a past merger event, possibly with an ultra-faint dwarf galaxy}. The dwarf-dwarf mergers can also leave discernible traces in their gas-phase metallicity. For instance, \cite{2008Michel-Dansac} showed that in a large sample of dwarf galaxy pairs with strong signs of interactions, the gas-phase metallicity is 0.2 dex higher than in their non-interacting counterparts.  

Observed morphological asymmetries and disturbances, such as shell-like features and tidal tails around dwarf galaxies, serve as a telltale of their relatively recent interactions \citep[e.g.][]{2003Conselice,2010Klimentowski, 2017Paudel2}. For instance, \cite{2004Coleman} suggested a past merger as a possible source of the small shell-like feature around the Fornax dwarf galaxy \citep{2012Yozin}. \cite{2014Amorisco} showed that the kinematically detected stellar stream in Andromeda II can be a possible remnant of a past dwarf-dwarf merger. Similarly, peculiarities in kinematics and chemical properties of other Local Group (LG) dwarf galaxies (such as Carina and Sculptor) are better explained as footprints of dwarf-dwarf encounters \citep{2004Tolstoy, 2012Venn}.

\begin{figure*}
\includegraphics[scale=0.8]{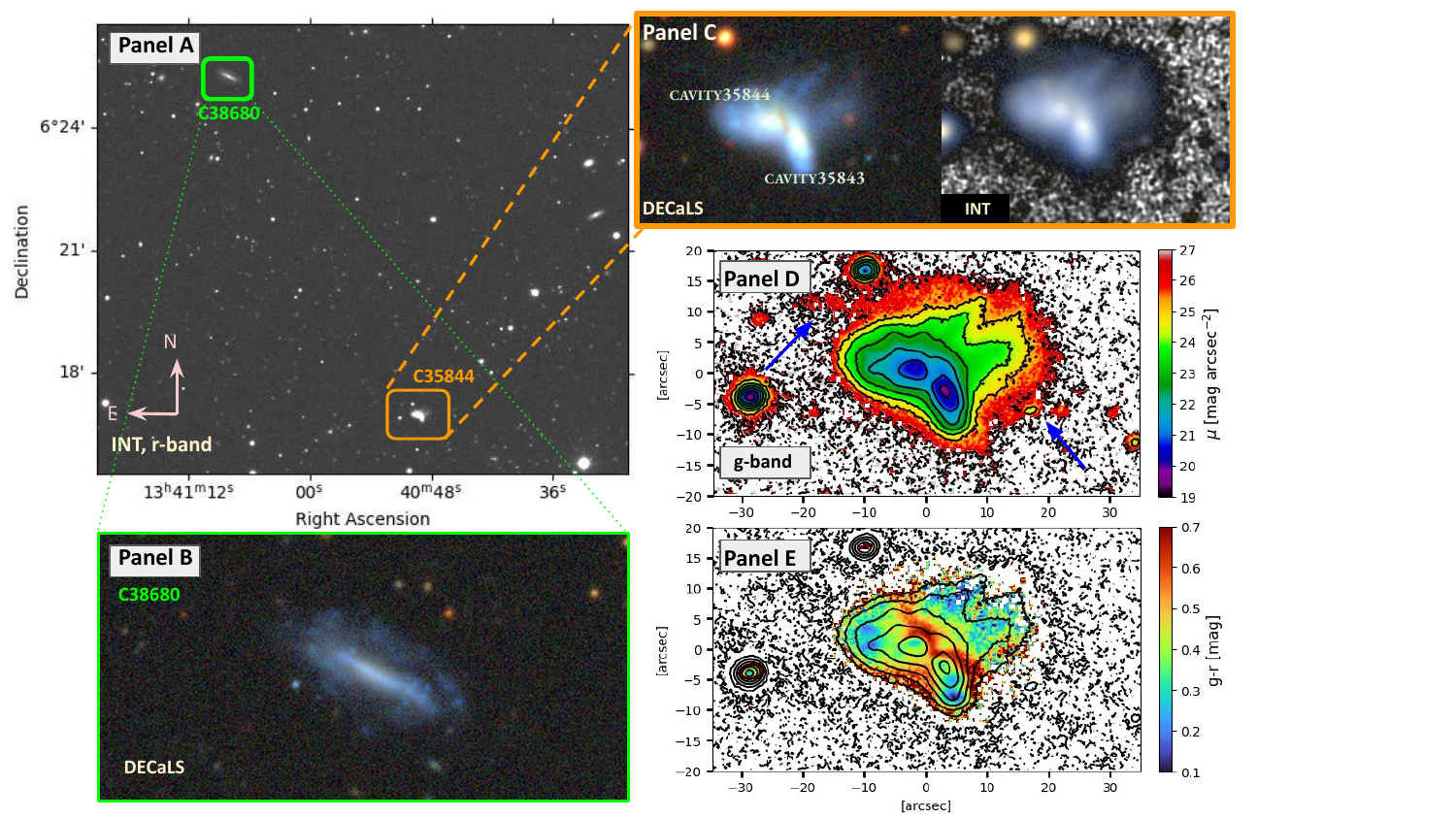}
\caption{Interacting pair of CAVITY35843, 35844, and their neighbour. \textit{Panel A:} Isaac Newton Telescope (INT) deep optical image of CAVITY35843 and CAVITY35844 (marked with an orange box) and CAVITY38680 (marked with a green box) in SDSS \textit{r}-band filter. \textit{Panel B:} Zoom-in view of CAVITY38680 using DECaLS data. This dwarf galaxy is located at a projected distance of 340 kpc from CAVITY35844. \textit{Panel C:} Zoom-in view of CAVITY35843 and CAVITY35844 (marked on the image) from DECaLS survey (left-hand panel) and INT \textit{g+r} image (right-hand panel). In both images, the two bright cores of the two interacting dwarfs and the tidal tails are visible. A fine dust feature in the pair central region is better resolved in the DECaLS image, while LSB features (with $\mu$ $>$ 25.5 mag  arcsec$^{-2}$, marked with arrows {in Panel D}) around the interacting pair are better visible in the INT colour image. \textit{Panel D:} Surface brightness map of the interacting pair using the \textit{g}-band INT image. Contours from inside out indicate regions with surface brightness from 20 to 27 mag  arcsec$^{-2}$ and with steps of 1 mag arcsec$^{-2}$. \textit{Panel E:} \textit{g-r} colour map of the interacting pair, constructed using INT data. Contours are the same as in Panel D. In all panels, north is up, and east is left.}
\label{Fig1}
\end{figure*}

Detection of mergers among dwarfs extends beyond the LG \citep[e.g.][]{2017Paudel1,2023ApJ...944..160M,2023A&A...671L...7R}. For instance, \cite{2012Graham} proposed that the merger of two dwarf galaxies featuring disc-like substructures could account for the rectangular shape of LEDA074886. \cite{2017Paudel2} investigated the merger origin of shell-like features in Virgo early-type dwarf galaxies. Subsequently, in \cite{2018Paudel}, they shed light on the prevalence of dwarf-dwarf mergers by presenting an extensive catalogue of nearby interacting and merging dwarf galaxies up to z~$\sim$~0.02, {but no information regarding the large-scale environment of this sample was given.} \cite{2021Kimbro} highlights the $\sim$ 10 kpc string of massive star clusters, with stellar populations younger than 10 Myr, bridging the northern and southern components of Mrk709 and suggests that it is possibly due to the early stages of a dwarf-dwarf galaxy merger.

Most of the dwarf-dwarf mergers investigated in the literature thus far are mainly located in the high-density regions (i.e. in galaxy groups and outskirts of clusters), where environmental effects are vigorous and non-negligible. In these cases, dwarf merging pairs are not only transformed due to the merging and close interaction with each other, but also experience tidal forces exerted by their host halo as well as hydrodynamical interactions with the surrounding intergalactic medium \citep[e.g.][]{2014Boselli}. Therefore, in such cases, disentangling between other environmental transformations and those solely induced by the merger event is not straightforward. One way to deepen our comprehension of dwarf galaxies' growth and transformation, specifically through mergers, is to investigate dwarf-dwarf mergers in {isolation \citep[e.g.][]{2020Kado-Fong, 2023Grajales-Medina, 2024A&A...686A.151C, 2024A&A...691A..82C, 2025arXiv250210078V}} and in pristine environments, such as cosmic voids \citep{2016Annibali}. Voids represent the most under-dense regions of the cosmic web \citep[e.g.][]{2001Peebles,2011Kreckel}, and they span large volumes \citep[i.e. the average size of 35 h$^{-1}$ Mpc;][]{2011Weygaert}. Within these vast expanses, galaxies are less disturbed and modified by the complex environmental processes in the galaxy groups and clusters. 

Few theoretical studies have focused on galaxy mergers in voids, revealing that the mean number of mergers does not depend significantly on the void environment \citep[e.g.][]{2022Rosas-Guevara}. The only difference reported so far between mergers in voids and those in clusters or groups is that mergers in voids tend to have occurred more recently (i.e. $<$ 2 Gyr), particularly for galaxies with log(M$_{\star}$/M$_{\odot}$) < 9.5 \citep{2024MNRAS.528.2822R}. From observations, several studies have also reported the presence of dwarf galaxies in voids with unusual features in their morphologies, which could be indicative of past mergers \citep[e.g.][]{2011Kreckel}. Interactions and mergers between small groups of galaxies have also been reported in some voids by a few studies \citep[e.g.][]{2013Beygu}.

In this study, we report the serendipitous detection of a gas-rich merging dwarf pair {located at the central regions (i.e. at R~$=$~0.13 of void effective radius)} of a cosmic void region that has been observed as part of the Calar Alto Void Integral-field Treasury Survey (CAVITY\footnote{\url{https://cavity.caha.es/}}) project \citep{2024A&A...689A.213P}. This legacy integral-field unit (IFU) survey aims to investigate the formation and evolution of nearby (0.005 $<$ z $<$ 0.050) void galaxies using their spatially resolved kinematics and stellar population properties \citep[e.g.][]{2023A&A...680A.111D,2024Conrado}.
The stellar mass ratio of galaxies in this merging pair, their local and large-scale environments, and their orientation with respect to the line of sight make it a great candidate for better understanding details of merger events in low-mass galaxies. To this end, we investigate spatially resolved (on kpc scales) ionised gas kinematics, gas-phase metallicity, and SFR of this isolated merging pair and compare our results with their other counterparts in denser environments than voids. The results of this study can shed light on the crucial role of mergers in shaping the present-day properties of dwarf galaxies and better distinguish between the footprints of environmental effects and those stemming from merger events in the mass assembly of dwarf galaxies.

This paper is organised as follows: In Section \ref{Data}, we introduce the IFU and deep photometric data sets used for the analysis. In Section \ref{Analysis}, we describe the methods we utilised for deriving kinematics and ionised gas emission. In Section \ref{Results}, we explain the results, and in Section \ref{Discussion}, we discuss them by comparing the interacting pair with its other counterparts from different environments in the literature. We summarise the key points of this work in Section \ref{Summary}. In this work, we assume a flat $\Lambda$CDM
cosmology with H$_{0}$=69.6 [km~s$^{-1}$ Mpc$^{-1}$], $\Omega_{\rm M}$ = 0.286, and $\Omega_{\rm vac}$ = 0.714.

\section{Data}\label{Data}
 The detected gas-rich merging pair consists of two dwarf galaxies, {designated as CAVITY35843, with stellar mass of log(M$_{\star}$/M$_{\odot}$) = 8.86 \citep[taken from the MPA-JHU catalogue;][]{2004Tremonti,2004Brinchmann}, and CAVITY35844 in the CAVITY survey}. In Table \ref{Pair_properties}, we summarised the main physical properties of these galaxies, {as reported in the literature}, including their CAVITY ID, right ascension (RA) and declination (DEC), redshift, the effective radius in \textit{r}-band, absolute \textit{r}-band magnitude, inclination, and the stellar mass. References for each value are provided in the table caption. In what follows, we describe the data used in this study.
 
\subsection{IFU data}\label{Data1}
The interacting pair was observed for a total exposure time {of 5400~s divided into three dithering positions, each with two exposures of 900~s,} using the PMAS/PPAK-IFU spectrograph at the 3.5 meter telescope of the Calar Alto Observatory (CAHA) as part of the CAVITY project and under the survey ID of CAVITY35844 \citep{2024A&A...689A.213P}. Observations were carried out using the V500 grating of the IFU instrument, which covers the optical wavelength range from 3750 to 7500\,\AA\ with a resolving power of  R $\sim$ 850 at 5000\,\AA\ and an average full width at half maximum (FWHM) of $\sim$ 6.0\,\AA. The instrument has a field-of-view (FOV) of 74 $\times$ 64 arcsec$^{2}$, with a spatial sampling of 1 arcsec, that provides adequate coverage of both interacting pairs and their tidal tail (see Fig.~\ref{Fig1}). {
Due to the small size of the galaxy pair, it occupies only the central region of the PMAS FOV and is therefore unaffected by vignetting effects.} Data reduction and construction of the final calibrated data cube were carried out through the CAVITY standard data reduction pipeline. In this process, along with other corrections and calibrations, the datacube has also been corrected for the Galactic extinction. For more details on the CAVITY data reduction pipeline, see \cite{2024arXiv241008265G}.

\setlength{\tabcolsep}{10pt}
\begin{table*}
\caption{\label{Pair_properties} {Properties of the merging pair as reported in the literature.}}
\centering
\begin{tabular}{c c c c c c c c}
\hline
CAVITY ID      & RA (J2000)$^{\rm a}$    &   DEC (J2000)$^{\rm a}$  &  $z$$^{\rm a}$  & \textit{\rm R}$_{\rm e}^{\rm b}$  & \textit{\rm M}$_{\rm r}^{\rm a}$  & inclination$^{\rm a}$ & {log(M$_{\star}$/M$_{\odot}$)$^{\rm c}$} \\
     &  [deg.] & [deg.] &  & [arcsec] & [mag] & [degree]\\
\hline
\hline
CAVITY35843 & 205.2043 & 6.282 & 0.029 & 5.48 &  -17.46  & 67.37 &  8.86\\
CAVITY35844 & 205.2057 & 6.283 & 0.029 & 9.87 &  -18.38 &  52.63 &  --- \\
\hline
CAVITY38680 & 205.2837 & 6.420 & 0.029 & 13.14 &  -17.90 &  75.51 & 8.80\\
\hline 
\end{tabular}\\
Properties of the merging system and its neighbour. a:\,\cite{2012Pan}, 
b:\,\cite{2015Alam},
{c: The median estimate of the total stellar mass probability density function (PDF) reported by \cite{2003Kauffmann}, using model photometry by \cite{2007Salim}}.
\end{table*}

\subsection{CAVITY+ deep optical imaging}\label{Data21}
{For our analysis, we took advantage of the deep optical imaging carried out with the 2.54\,m Isaac Newton Telescope (INT), as part of the CAVITY+ survey \citep{2024A&A...689A.213P}.} This survey uses the Wide Field Camera (WFC), mounted on the INT primary focus, that approximately covers a FOV of 34' $\times$ 34' with a pixel scale of 0.33 arcsec pix$^{-1}$. In Fig.~\ref{Fig1} Panel~A, we show CAVITY35843 and 35844 (marked with an orange box) along with CAVITY38680 (their only neighbouring galaxy within a projected search area of 1.5 Mpc, marked with a green box, {see Table~\ref{Pair_properties}}) in the reduced and calibrated INT images taken in the \textit{r}-band filter.  For details on the CAVITY+ deep optical imaging survey and data reduction, see \cite{2024A&A...689A.213P}.

All three galaxies shown in Panel A of Fig.~\ref{Fig1}  belong to the same void. CAVITY38680 is located at a distance of 0.343 Mpc to the interacting pair, with a difference in line-of-sight velocity of 219.4 km~s$^{-1}$ with respect to the pair. {The distance and relative velocity are computed based on the RA, DEC, and redshift of these galaxies (see Table~\ref{Pair_properties}) following the methodology of \cite{2015A&A...578A.110A}}. A zoom-in view of CAVITY38680 is shown in Panel B, using data from the Dark Energy Camera Legacy Survey \citep[DECaLS,][]{2016Blum}. CAVITY38680 is another dwarf galaxy with log(M$_{\rm \star}$/M$_{\rm \odot}$)~=~8.8 and is actively star-forming (log(SFR) = -0.51 [M$_\odot \rm yr^{-1}$]) \citep[][]{2004Tremonti,2004Brinchmann}. This dwarf galaxy has a thin disc {(or a large bar)} surrounded by irregular features in its outskirts. {According to \cite{2015A&A...578A.110A}, CAVITY38680 is not physically bound with CAVITY35843 and 35844, but it belongs to the same large-scale structure environment. This means that the neighbour galaxy in this case does not affect the evolution of the interaction, and the merger is located in a low-density local environment. }

\begin{figure*}
\centering
\includegraphics[width=\textwidth]{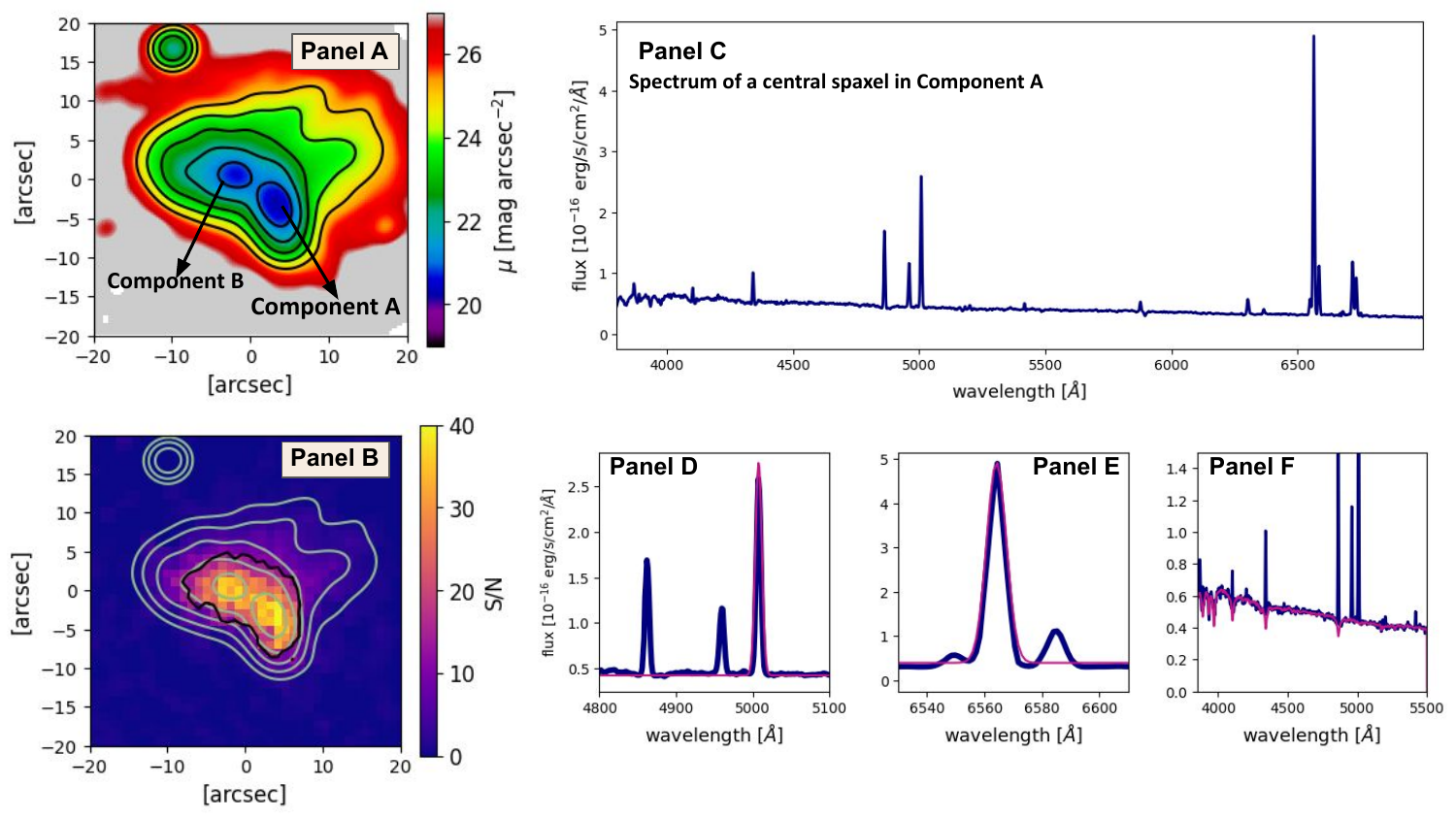}
\caption{Observed IFU data. \textit{Panel A:} \textit{g}-band surface brightness map of CAVITY35843 and 35844, same as panel D of Fig.~\ref{Fig1} but convolved to match the pixel scale of PPAK IFU data.  \textit{Panel B:} Signal-to-noise map of the PPAK data with the same contours as in Panel A. {The black contour marks the S/N = 15. }\textit{Panel C:} Example of PPAK spectrum, {in the rest frame}, from a central spaxel in component A. \textit{Panels D and E:} {Examples of single-line Gaussian fits for} [OIII]$\lambda$5007 and H$\rm \alpha$ emission lines, respectively. The best fit is shown in pink. \textit{Panel F:} Fit over {the stellar continuum of} the same spectrum shown in Panel C using pPXF (see Sec.~3 for details).} 
\label{Fig2}
\end{figure*}

In Panel C of Fig.~\ref{Fig1}, we show a zoom-in view of CAVITY35843 and 35844 from DECaLS (left-hand Panel) and INT \textit{g+r} images (right-hand Panel). {While the data coming from INT have much higher surface brightnesses \citep[30.0 and 29.0~mag~arcsec$^{-2}$ in \textit{g} and \textit{r} bands respectively, measured in 3 sigma, 10$\times$10 arcsec boxes;][]{2020A&A...644A..42R}, their seeing is mediocre (1.9 and 2.0 arcsec FWHM). To address this, we utilise DECAM instrument data\footnote{Dark Energy Camera, \url{https://noirlab.edu/science/programs/ctio/instruments/Dark-Energy-Camera}}, that have slightly lower surface brightness limits (approximately 1~mag~arcsec$^{-2}$ less \citealt{2023A&A...671A.141M}), but offer superior seeing conditions, with FWHM values of 1.18 and 1.29 arcsec in the \textit{r} and \textit{g} bands, respectively. }Both data show two bright cores, representing the core of each interacting dwarf component, and a faint tidal tail heading toward the north-west. In Panel D of the same figure, we show the surface brightness map of this interacting pair using \textit{g}-band INT images. The two bright cores of this interacting pair have similar surface brightness 19.5$<$\,$\mu$\,[mag\,arcsec$^{-2}]$\,$<$20.5 in \textit{g}-band. The tidal tail is 4 to 5 mag fainter than the two cores. Additionally, we could trace two low surface brightness (LSB) features ($\mu$ $>$ 25 mag arcsec$^{-2}$, marked with arrows {in panel D}) in the north-east and south-west of the interacting system, that are possibly the results of the ongoing interaction between these two dwarf galaxies. These LSB features are not traced in any other optical data available in the literature (including DECaLS presented here).

In Panel E of Fig.~\ref{Fig1}, we show the \textit{g-r} colour map of CAVITY35843 and 35844, computed using INT deep images in SDSS \textit{g} and \textit{r} filters. Blue regions with 0.10$<$ \textit{g-r}[mag]$<$0.25, possibly dominated by star formation, are seen in the tidal tail and the galaxy's east and west parts. Additionally, we trace a curved strip-like feature, dominated by red colours (\textit{g-r} $>$ 0.6 mag), that separates the two bright cores of the interacting pair and their tidal tails. Compared with the DECaLS colour image in Panel C, one can notice that this red curved strip traces the distribution of dust. In the southern part of the interacting pair, we also observe a distribution of redder light. However, no dust lanes were visually detectable in these regions in the DECaLS data; thus, they may trace the presence of older or metal-richer stellar populations. 

\begin{figure*}
\centering
\includegraphics[width=\textwidth]{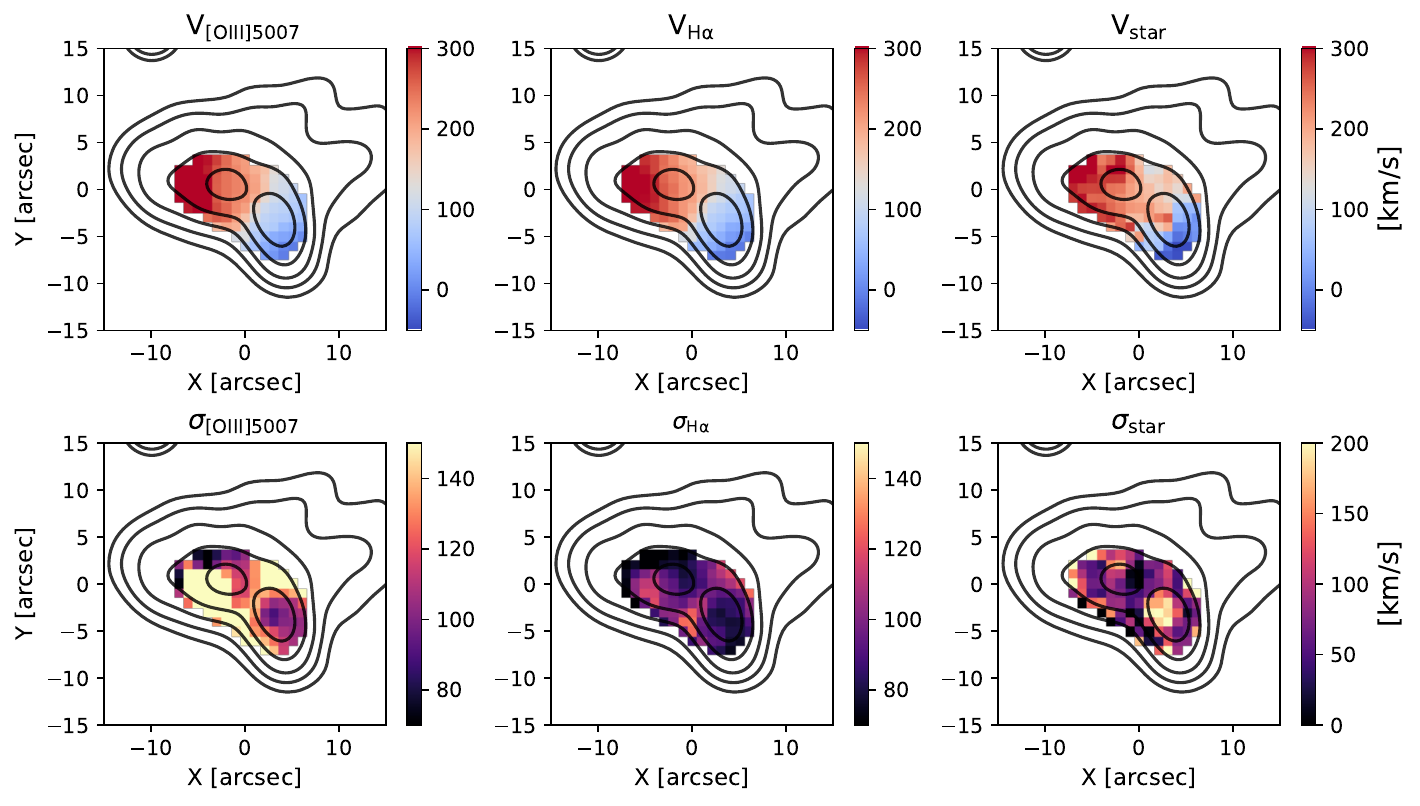}
\caption{Gas and stellar kinematic maps of the merging pair. In the left-hand and middle panels, gas radial velocity (top) and velocity dispersion (bottom) maps derived from [OIII]$\lambda$5007 and H$\rm \alpha$ emission lines are shown, respectively. In the right-hand panel stellar radial velocity (top) and velocity dispersion (bottom) maps are shown. In all panels, only spaxels with S/N  $>$ 15 are shown. Contours are similar to Fig.~\ref{Fig2}.}
\label{CAVITY35844_Kinematics}
\end{figure*}

In panels A and B of Fig.~\ref {Fig2}, we show the surface brightness (from the INT \textit{g}-band image) and signal-to-noise ratio (S/N, extracted from PPAK data) maps of the merging pair, respectively. Here, we convolved the INT surface brightness map to match the spatial resolution of the PPAK data {(i.e. with a pixel size of 1 arcsec corresponding to the spatial resolution of 2.5 arcsec)}. The S/N is defined on the continuum level of each spaxel’s spectrum and within the 4590 to 4610 \AA\, wavelength range. As marked on the surface brightness map and to ease the discussion hereafter, we refer to the brightest interacting galaxy (i.e. CAVITY35843) as component A; the other (i.e. CAVITY35844) is referred to as component B. In our analysis, we excluded spaxels of the PPAK data with S/N $<$ 15 (except in the analysis in Section \ref{Result2}) to attain reliable measurements. This mainly excludes the outskirts and faint tidal tail of the merging pair.

\section{Analysis}\label{Analysis}

To derive the gas and stellar kinematics and measure fluxes of emission lines, we first masked out foreground stars and background galaxies in the FOV of IFU data, and utilised the publicly available full spectral fitting penalised pixel-fitting algorithm \citep[pPXF;][]{2004Cappellari, 2017Cappellari}. To estimate the stellar kinematics, pPXF requires a set of sample stellar models for which we used the \cite{2010Vazdekis,2016Vazdekis} single stellar population (SSP) models based on the MILES stellar library \citep{2006MNRAS.371..703S,2007Cenarro,2011A&A...532A..95F} with spectral FWHM of $\sim$\,2.51~\AA. SSP models used here {are based on BASTI isochrones \citep{2004Pietrinferni} and a bi-modal initial mass function with slope of 1.3 \citep{1996Vazdekis}} and cover the age range of 0.03 to 14.00 Gyr, the metallicity range of -2.27 to 0.40 dex, {with} the base [$\alpha$/Fe].

We derived the kinematics of both the gas and stellar components and measured the gas fluxes through two iterations of pPXF over each spaxel {(no spatial binning was performed)}.  In the first iteration, we estimated the noise of each spectrum using residuals of the pPXF fit and then clip outliers (on 3$\sigma$ level) for the second iteration. The results are extracted from the latter iteration using 14th-order additive Legendre polynomials. The spectrum of a central spaxel in component A is shown in panel C of Fig.~\ref{Fig2}, and an example of Gaussian single fits over the gas emission lines is illustrated in panels D and E. The derived radial velocities and velocity dispersions ($\sigma$) of the emission lines through this procedure are shown in Fig.~\ref{CAVITY35844_Kinematics}. 

{In panel F of Fig.~\ref{Fig2}, we present the stellar component of the pPXF fit applied to the spectrum of a central spaxel, as an example.} It is evident from the fit that this system is heavily dominated by gas emission lines whose presence makes analysis of the underlying stellar component challenging and unreliable. Binning the data, though not advisable in this case as it can obscure valuable spatial information, also did not improve the fits over the stellar component. {Therefore, in this study, we refrain from discussing the stellar population properties of the pair.}

\section{Results}\label{Results}
\subsection{Kinematics}\label{Results1}

In the top row of Fig.~\ref{CAVITY35844_Kinematics}, we present the radial velocity maps derived from gas and stellar components. Here we show the radial velocity of [OIII]$\lambda$5007 emission line, representing the forbidden lines (left-hand panel), and the H${\rm \alpha}$ emission line, representative of the Balmer series in this system (middle panel). The radial velocity map derived from the stellar component is shown on the right-hand panel. All three maps show a lopsided rotation in both the stellar and gas components. The similar range of stellar and gas radial velocity (-50 $<$ V$_{\rm rot}$ [\rm km~s$^{-1}$] $<$ 300) and their spatial distribution indicate that both gas and stellar components are co-rotating in a common structure. Part of the lopsidedness observed in these maps is due to the relative velocity between component A and component B ($\sim$ 100 km~s$^{-1}$). {Separating these components to account for the velocity difference allows the detection of rotation in both merging dwarf galaxies (see Appendix \ref{AppA}).}

In the bottom row of the Fig.~\ref{CAVITY35844_Kinematics}, we show the velocity dispersion of the gas components (i.e. $\sigma_{\rm gas}$) and the stellar component ($\sigma_{\rm star}$). The $\sigma_{\rm gas}$ maps show fluctuations in values that could be expected in an entangled and complicated interacting system such as here. In particular, regions around both distinct bright central components of this interacting system have the lowest $\sigma_{\rm gas}$ of $\sim$ 70$\pm$5 km~s$^{-1}$ and $\sim$ 150$\pm$25 km~s$^{-1}$ measured from H$\alpha$ and [OIII]$\lambda$5007 emission lines, respectively. Both maps show that toward the system's outskirts and the faint tail, the $\sigma_{\rm gas}$ values become larger {($>$ 130 km~s$^{-1}$)} with average errors of 17 km~s$^{-1}$ and 25 km~s$^{-1}$ for values derived from H$\alpha$ and [OIII]$\lambda$5007, respectively. Both maps show the highest $\sigma_{\rm gas}$ values in a small region on the left-hand side, reaching $\sim$ 300 $\pm$ 175 km~s$^{-1}$.  {The detected emission line was best fit with a single Gaussian, and visually, we could not identify any contributions from multiple components in the emission line profiles for any spaxel with S/N > 15 in this system (see Appendix \ref{AppA}). }Hence, this fluctuation can partly be due to lower S/N in the system’s outskirts. We also believe that the unresolved contributions of both interacting components, which could not be appropriately disentangled with the given data resolution, are partly responsible for larger $\sigma_{\rm gas}$ values in the regions between two bright central components and the outskirts, as they can alter the width of gas emission lines. {However, it is important to note that the spectral resolution of the PPAK V500 mode (R$=$ 850 at 5000 \AA) is significantly lower than the variations and values measured in these maps. Therefore, the results should be interpreted with caution.}

We noticed that through the whole extension of this system, the $\sigma$ measured from the [OIII]$\lambda$5007 line (with median of 116$\pm$18 km~s$^{-1}$) is larger than $\sigma_{\rm gas}$ derived using the H$\alpha$ (with median of 82$\pm$7 km~s$^{-1}$). However, this difference (at 2$\sigma$ level) is not significant and might be due to, for example, the presence of multiple interstellar medium components heated by different mechanisms (such as star formation) that are unresolved in the present data.

The $\sigma_{\rm star}$ map is more complicated as it shows higher values in the central regions of component A and much lower values, with a difference of $>$ 100 km~s$^{-1}$ in component B. Since CAVITY35843 and CAVITY35844 are in an entangled system with prominent emission lines, pPXF could not retrieve a reliable stellar $\sigma$ for most spaxels. Hence, we do not comment further on the $\sigma_{\rm star}$ measurements.

\cite{2022Bik} derived the gas kinematics of the component A using near-infrared integral-field K-band spectroscopy data. The lopsidedness and the range of gas rotational velocity measured for this component match well with the range that \cite{2022Bik} have reported ($-$50 $<$V [km~s$^{-1}$]$<$75). They also measured an average $\sigma$ gas of 54.4 $\pm$ 9.6 km~s$^{-1}$, consistent with the range of values we measured for this component using the H$\alpha$ emission line. Moreover, \cite{2022Bik} derived stellar kinematics of component A using CO absorption lines and reported the rotation range of $\pm$23 km~s$^{-1}$, and we found this range to be consistent with what was obtained from PPAK data for both the gas and stellar components in the optical range (see Fig. \ref{CAVITY35844_Kinematics}).

\subsection{Separating the kinematics of interacting components}\label{Result2}
\begin{figure}
    \centering
    \includegraphics[scale=0.64]{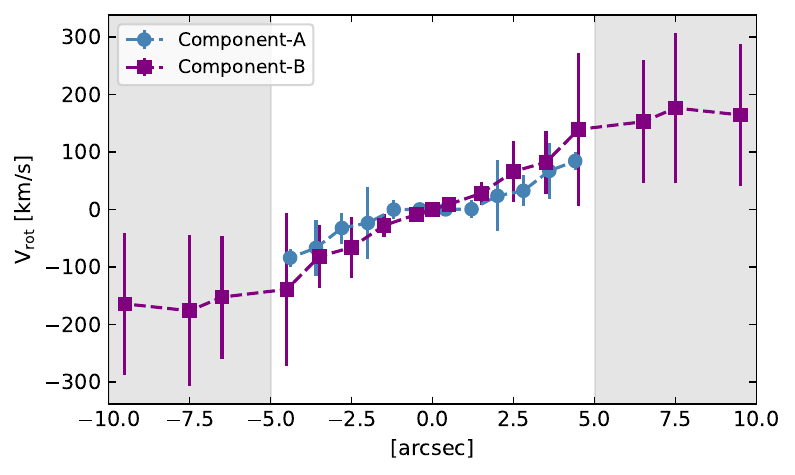}
    \includegraphics[scale=0.565]{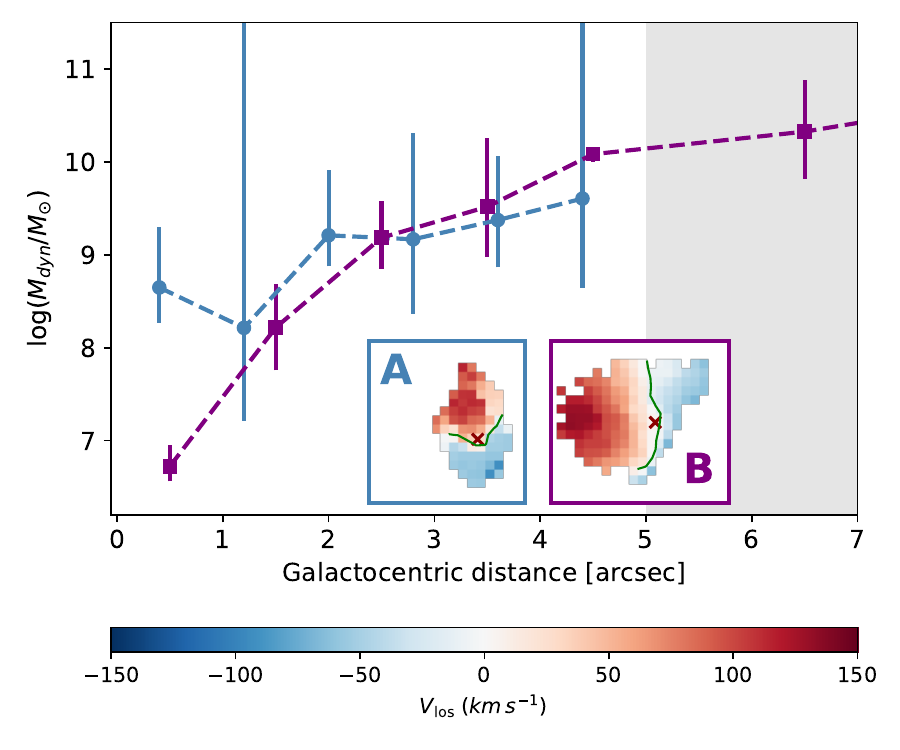}
    \caption{Rotating disks in the merging dwarfs. \textit{Top panel:} Rotational velocity of each separated component derived using ${\tt ^{3D}BAROLO}$. Both components show a similar radial gradient of rotational velocity by tracing the H${\alpha}$ emission line. We note that results beyond $\sim$ 5 arcsec are excluded from our discussion as they are derived from fits over spaxels with S/N $<$ 15. \textit{Bottom panel:} Enclosed dynamical mass of each component calculated at a given radius using the rotation curves estimated by ${\tt ^{3D}BAROLO}$, assuming spherical symmetry. {Each mass is the enclosed at a given radius.} The rotational velocity map of each component is shown inside boxes. In each map, the assumed {kinematical} centre of the galaxies is marked. The green line represents the V$_{\rm rot}$ = 0 km~s$^{-1}$ isoline of each velocity map.}
    \label{Fig4}
\end{figure}

Although components A and B are interacting, their central bright nuclei are still distinct (see Fig.~\ref{Fig1}). Also, in the kinematic maps of Fig.~\ref{CAVITY35844_Kinematics}, these two regions seem to be separated with a relative velocity of $\sim$ 100 km~s$^{-1}$ and a fine line of high $\sigma$ values that, as discussed before, could be the result of unresolved contributions from both of the interacting components in each spaxel's spectrum. To take advantage of their unique configuration, we separated the components and investigated their kinematics separately.

For this purpose, we used ${\tt ^{3D}BAROLO}$ on the H$\alpha$ emission line extracted from the IFU datacubes. ${\tt ^{3D}BAROLO}$ is a publicly available software designed to measure rotation curves of galaxies using their gas emission lines \citep{2015Teodoro}. This code builds 3D tilted-ring models for galactic discs, assuming that the observed wavelength shift in all the emission lines within the disc and in each ring is governed by the galaxy's rotation and that the V$_{\rm rot}$ is fixed per radius \citep{1974Rogstad}. Hence, ${\tt ^{3D}BAROLO}$ obtains the best geometrical and kinematical fit parameters by generating multiple models using the Monte Carlo approach and comparing them with the data. 
\begin{figure*}
\centering
\includegraphics[width=\textwidth]{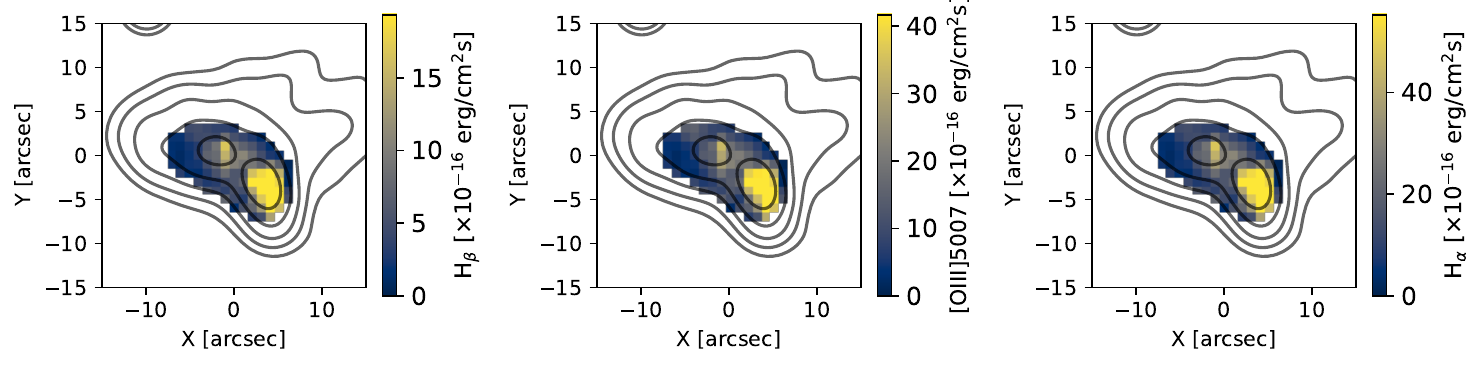}
\caption{Emission line maps. From left to right, maps of H$\beta$, [OIII]$\lambda$5007, and H$\alpha$ fluxes. All the emission lines shown here are corrected for optical dust attenuation, and contours are the same as in Fig. \ref{Fig2}. {In all maps, only spaxels with S/N$>$15 are shown where we considered the same S/N definition, computed in the wavelength range from 4590 to 4610 \AA\,}.}
\label{CAVITY35844_gas_maps}
\end{figure*}

To find the best masking scheme for separating the merging components, we ran several tests for different morphologies and configurations. In the end, we found that the best mask separates the two components alongside the high $\sigma_{\rm gas}$ line between the bright nuclei, as discussed earlier and shown in Fig.~\ref{CAVITY35844_Kinematics_AppendixA} of Appendix~\ref{AppA}. Then, with the ${\tt ^{3D}BAROLO}$, we individually analysed each interacting component.

In this work, we derived the rotation curves of each component solely from the H$\alpha$ emission line (i.e. isolating the H$\alpha$ emission line within a $\pm$100 \AA\, window around the observed wavelength), as it is the most prominent line in our dataset. Also, based on our results from Fig.~\ref{CAVITY35844_Kinematics}, we do not expect significant differences in the rotational velocity derived from the Balmer and forbidden lines. Before modelling, we also removed the stellar continuum from the spectra using a first-degree polynomial fit on both sides of the [NII] emission lines{\footnote{We used \texttt{specutils.fitting.fit\_generic\_continuum()} function with a linear model to fit the stellar continuum.}. 

To model the tilted rings, ${\tt ^{3D}BAROLO}$ requires estimates for several initial parameters, such as the central coordinates of each disc, galaxy's inclination, position angle, redshift, thickness of the disc, velocity dispersion, and rotational velocity. The first three parameters were estimated via an isophote analysis of the galaxies using Astropy's \texttt{photutils.aperture} module, while we assumed a uniform {azimuthal} width of 1 arcsec for the discs. {Both merging components were assumed to be disky, primarily based on the detected rotational kinematics (see Appendix \ref{AppA}) and the presence of strong gas emission lines, both characteristic features of late-type disky dwarf galaxies.} Initial velocity guesses were taken from the results shown in Fig.~\ref{CAVITY35844_Kinematics}.

We ran ${\tt ^{3D}BAROLO}$ several times over each interacting component, manually fine-tuning the initial parameters estimates each time to better model the kinematic and distribution of the gas component. This was needed to retrieve reliable values due to the limited spatial resolution of the data. {All the input parameters used for ${\tt ^{3D}BAROLO}$'s runs are available in the Appendix \ref{AppB}.}

In Fig.{~}\ref{Fig4} we present the results of the ${\tt ^{3D}BAROLO}$ analysis. On the top panel of this figure, we show the rotational velocity curves for each component. They exhibit a similar range of rotation up to $\sim$ 5 arcsec from the galaxy centre. Beyond this distance, our data only has spatial coverage for component B. We note that, as indicated by the error bars, the uncertainty for values measured beyond this distance is larger, and this is primarily because the ${\tt ^{3D}BAROLO}$ analysis was conducted without considering any criteria on spaxels S/N. {We do not consider values beyond 5 arcsec in this study, since they are derived based on spaxels with inadequate S/N.}

In the bottom panel of Fig. \ref{Fig4} we show the dynamical mass curve of each merging component. This mass is estimated from the rotational velocity derived using ${\tt ^{3D}BAROLO}$ and assuming spherical symmetry for both components {as a first order approximation}:
\begin{equation}
M_{\text{dyn}}(R) = V(R)^2\frac{R}{G},
\end{equation}
where $V(R)$ is the rotation velocity as a function of the radius, $R$, and $G$ is the Newtonian constant of gravitation. The curves shown in Fig.~\ref{Fig4} indicate that both components have a similar range of dynamical masses up to a galactocentric distance of 5 arcsec, that corresponds to 2.9 kpc. {The dynamical mass of each galaxy is, within this range, $\log \left(M_{\rm dyn, A}/M_{\odot}\right)=9.6\pm2.2$ and $\log \left(M_{\rm dyn, B}/M_{\odot}\right)=10.1 \pm 0.1$, for component A and B respectively. We note that these values represent lower limits to the dynamical mass, as the data used here do not cover the full extent of the emission.} {Additionally, beam smearing, particularly in cases with low angular resolution (i.e. 2.5 arcseconds in this study) can affect the rotation curves, decreasing their slopes, that may also lower the dynamical masses estimated here.} The larger value estimated for the dynamical mass of component A in the very central region may be due to an offset in the estimation of the galaxy's centre {but it is not ultimately reliable physically because the following value has large errorbars}. We attempted multiple fits with varying centre estimates, but none resulted in a  model that adjusts the galaxy's rotation better for this region.

{The use of ${\tt ^{3D}BAROLO}$ for these data is fundamentally risky, as it relies on the assumption that each interacting component exhibits some level of symmetry. Furthermore, the kinematics we are modelling are likely influenced by the conditions under which the two galaxies interacted and the current stage of the merger. While this cannot be thoroughly evaluated with the available data, it undoubtedly contributes to the uncertainty in the results. Additionally, the low resolution of data used here, especially when both galaxies are separated, adds to the uncertainties. Despite these limitations, this test allowed us to confirm the hypothesis of a nearly 1:1 merger within the error bars intrinsic to the nature of the data.}

\setlength{\tabcolsep}{20.pt}\label{Results3}
\begin{table*}
\caption{\label{Gas_properties} Integrated emission line fluxes.}
\centering
\begin{tabular}{c c c c}
\hline
Emission line   &  Observed total flux & Component A & Component B   \\
                &  $\times$ 10$^{-16}$ [erg/s/cm$^2$] &  $\times$ 10$^{-16}$ [erg/s/cm$^2$] &  $\times$ 10$^{-16}$ [erg/s/cm$^2$] \\
\hline
\hline
H$\gamma$ & 462.2 $\pm$ 2.2 &  281.0 $\pm$ 2.0  & 168.6 $\pm$ 2.1\\
H$\beta$ & 960.6 $\pm$ 2.4 &  574.5 $\pm$ 2.4  & 337.3 $\pm$ 2.3\\
H$\alpha$ & 2747.4 $\pm$ 3.0 &   1643.0 $\pm$ 2.5 & 964.7 $\pm$ 2.7\\
$ \rm [OIII]$$\lambda$5007  & 2097.7 $\pm$ 3.1 &  1273.9 $\pm$ 2.9  & 736.6 $\pm$ 2.9\\
$ \rm [NII]$$\lambda$6583  &  662.4 $\pm$ 3.8  &  412.7 $\pm$ 3.3  & 207.5 $\pm$ 3.4\\
$ \rm [SII]$$\lambda$6716  &  510.5 $\pm$  3.1 &  3112.8 $\pm$ 3.5 & 172.7 $\pm$ 2.8\\
$ \rm [SII]$$\lambda$6731  &  357.3 $\pm$ 3.1  &  224.8  $\pm$ 2.7 & 117.0 $\pm$ 2.8\\
\hline

\end{tabular}\\
Measured integrated emission line fluxes. Columns list the name of the emission line, the fluxes measured over the integrated spectra of the entire merging system, and the fluxes for components A and B. All fluxes are corrected for intrinsic reddening. \\
\end{table*}

\subsection{Emission lines and ionisation sources}\label{Result3}

As explained in Section \ref{Analysis}, we measured the emission line fluxes for each spaxel with S/N $>$ 15 using the pPXF pipeline. These fluxes are computed after subtraction of the stellar continuum. We then corrected them for the intrinsic reddening using the \cite{1989Cardelli} extinction law. We assumed R$_{\rm v}$ = 3.1 and the Balmer decrement (H${\alpha}$/H${\beta}$) = 2.86 \citep{1984Osterbrock}. In Fig. \ref{CAVITY35844_gas_maps}, we present maps of extinction corrected emission line fluxes for [OIII]$\lambda$5007 and two Balmer lines, namely H${\rm \alpha}$ and H${\rm \beta}$.

{To facilitate a more accurate comparison with results in the literature, which are primarily based on long-slit spectra of other merging dwarf systems, we constructed integrated spectra for each merging component as well as the entire system. This was achieved by summing the spectral data over spatial elements (spaxels) with S/N $>$ 15. For each component, the same masking scheme described in Section \ref{Result2} was applied to create the integrated spectra (see also Appendix \ref{AppA}). Gas emission fluxes were then derived from these spectra using a strategy similar to that outlined in the beginning of Section \ref{Analysis}. The reddening-corrected emission line fluxes obtained from the integrated spectra are listed in Table~\ref{Gas_properties}. A comparison of these fluxes, as well as the maps in Fig.~\ref{CAVITY35844_gas_maps}, particularly for the Balmer lines, clearly indicates that component A is more gas-rich than component B.}}

\begin{figure*}
\centering
\includegraphics[width=\textwidth]{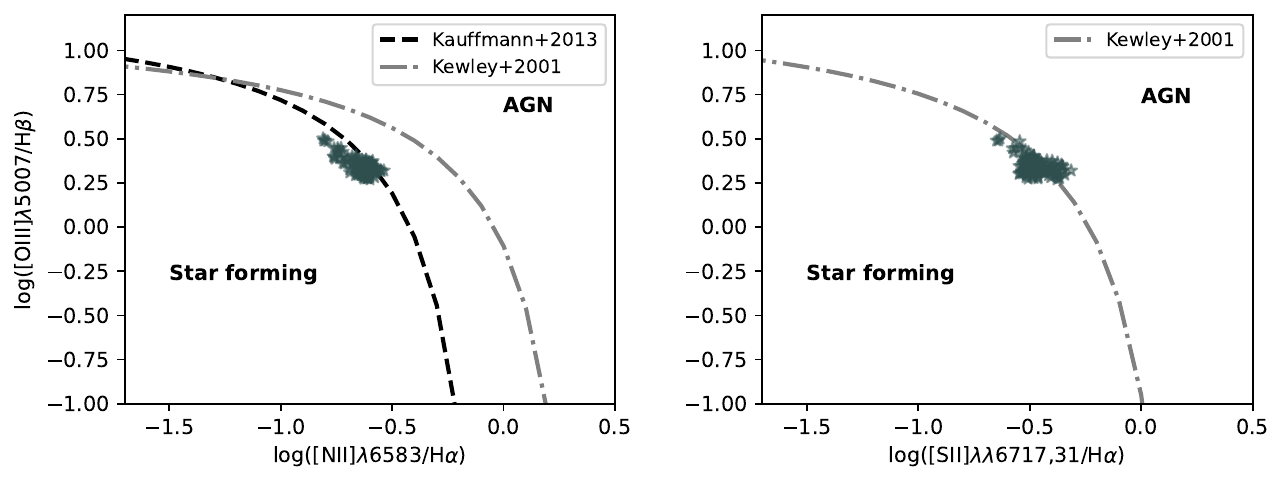}
\caption{Baldwin, Phillips
and Terlevich diagrams using flux measured for each spaxel with S/N $>$ 15 in both components of the merging system. In both panels, \protect\cite{2001Kewley} theoretical division is traced with a gray dashed-dot line. The empirical division of \protect\cite{2003Kauffmann} is also shown in the left-hand panel with a black dashed line. The intermediate zone between the empirical and theoretical divisions in the left-hand panel is known as the composite zone. }
\label{CAVITY35844_BPT}
\end{figure*}

To investigate the nature of the ionisation mechanisms present in this merging pair we utilised the `Baldwin, Phillips and Terlevich' (BPT) diagnostic diagram \citep[Fig.~\ref{CAVITY35844_BPT};][]{1981Baldwin}, with the combination of log([OIII]$\lambda$5007/H$\beta$) vs. log([NII]$\lambda$6583/H$\alpha$). This combination is commonly used to discern the ionisation source of emission lines in galaxies. {Emission lines positioned below the \cite{2003Kauffmann}’s empirical division (the black dashed line) are mainly ionised by star formation, and those above the \cite{2001Kewley}’s theoretical one (the gray dashed-dotted line) are AGN dominated.
} Determining the ionisation origin for galaxies (or spaxels, in this case) that fall within the intermediate zone between these two divisions (known as the composite zone) is more complex. Therefore, in the right-hand panel of Fig. \ref{CAVITY35844_BPT}, we also analysed the combination of log([OIII]$\lambda$5007/H$\beta$) vs. log([SII]$\lambda\lambda$6717,6731/H$\alpha$). As in previous sections, only spaxels with S/N $>$ 15 are included in both panels. Most spaxels are located within the star-forming zone, with a few in the AGN zone. The latter spaxels, situated mainly at the outskirts of the interacting system (see Appendix \ref{AppC}), are likely affected by low S/N.

{Additionally, we measured the equivalent width of the H$\alpha$ emission line, EW(H$\alpha$), following \cite{2011CidFernandes} to identify any possible weak AGN activity. Spaxels with S/N $>$ 15 in this merging system show a range of -0.73 $<$ log(EW(H$\alpha$)) $<$ 1.46 and -0.81 $<$ log([NII]$\lambda$6584/H$\alpha$) $<$ -0.53, confirming that AGN activity is not the ionisation source in this system.} From Fig. \ref{CAVITY35844_BPT} and the EW(H$\alpha$) measurements, it is evident that gas in both components of this system is predominantly ionised by radiation from massive, recently formed stars and possibly by the tidal shocks that are capable of elevating the low-ionisation lines such as the [NII] and [SII]. {This is further supported by the shallow radio spectral index of -0.63$\pm$0.09 reported by \cite{2018deGasperin} for this system, which is consistent with a mix of nonthermal radio emission from cosmic-ray electrons, supernova remnants, and thermal bremsstrahlung emission from ionised gas.} Thus, we can confidently exclude AGN as a potential ionisation source in this system.

{Using the stacked spectra of each component, we estimated the stellar mass through full spectral fitting (see Section~\ref{Analysis}). The derived stellar masses for each component, along with the total stellar mass of the system, are reported in Table~\ref{integrated_values}. These estimates support the findings of Section~\ref{Result2}, confirming that this system is a $\sim$ 1:1 merger. However, it is important to note that these stellar masses are derived from full spectral fitting of the stellar (continuum) component of the merging system. Given that the spectra are heavily dominated by emission lines and the complex, intertwined nature of the data, which makes precise separation of the merging components challenging, these values should be regarded as approximations and interpreted with caution.
}

\setlength{\tabcolsep}{25.pt}\label{Results3}
\begin{table*}
\caption{\label{integrated_values} Global properties measured from IFU data.}
\centering
\begin{tabular}{c c c c}
\hline
Parameter   &  Total system & Component A & Component B   \\
            &               & CAVITY35843 & CAVITY35844   \\
\hline
\hline
$\log\left(M_{\rm dyn}/M_{\odot}\right)$ & 10.22 & 9.60$\pm2.2$ &  10.10$\pm 0.1$\\
$\log\left(M_{\rm \star}/M_{\odot}\right)$ & 9.7 & 9.3 &  9.5\\
12+log(O/H) & 8.55 &  8.55  & 8.51\\
log(SFR [M$_\odot \rm yr^{-1}$]) & 0.61 & 0.38 & 0.15\\
log(sSFR [yr$^{-1}$]) & -9.1 & -8.92 & -9.35\\
\hline
\end{tabular}\\
Physical properties of merging components. The columns include the name of the measured parameter, along with the values obtained for the total system, component A, and component B. The reported dynamical mass is derived within 5 arcseconds of each component. The gas-phase metallicity is based on the empirical calibration of the N2 index \citep{2004Pettini}. The star formation rate (SFR) is calculated using the equation from \cite{1998Kennicutt}.  \\
\end{table*}

\subsection{Gas metallicity}\label{Results4}

We assessed the oxygen abundance (O/H) for spaxels with S/N~$>$~15, as a proxy for gas phase metallicity in this merging pair using the empirical calibration of the N${\rm 2}$ index as proposed by \cite{2004Pettini}:
\begin{equation}
\rm 12+log(O/H) = 8.90 + 0.57 \times N{\rm 2},
\end{equation}
where N${\rm 2}$ $\equiv$ log($\frac{[NII]\lambda6583}{H\alpha}$). The N${\rm 2}$ empirical calibrator is well suited for the analysis of moderate S/N data and is a better choice for low-mass systems such as CAVITY35843 and CAVITY35844 \citep[e.g.][]{2021Zurita}. A typical systematic uncertainty of $\pm$0.2 dex in metallicity derived using the N${\rm 2}$ index is expected according to \cite{2004Pettini}. In the top panel of Fig. \ref{CAVITY35844_dust_metal_SFR} we show the distribution of 12+log(O/H) values measured for each spaxel. {We also derived the global gas-phase metallicity from the integrated spectra described in Section \ref{Result3}, using the N${\rm 2}$ index. For the integrated spectra of both components, we obtained 12+log(O/H) = 8.55 dex, that matches the value measured from the integrated spectra of component A. Component B shows a slightly lower value of 12+log(O/H)~=~8.51~dex (see Table~\ref{integrated_values}).} It is evident from the Fig.~\ref{CAVITY35844_dust_metal_SFR} map and integrated values that, on average, both of the components have similar near-solar gas-phase metallicities, particularly in their central regions, ranging between 8.50$<$12+log(O/H)~[dex]$<$8.58, \citep[compared to the solar value of 12+log(O/H) = 8.66 dex;][]{2004Asplund}.

This map also shows a gradient in metallicity values between the central and outskirts of the system, ranging from $\sim$ 8.45 dex in the north-east to above 8.60 dex in the south-west. However, we do not discuss this gradient any further as the difference between the outskirts and central regions of the interacting pair ($\sim$~0.15 dex) is within the systematic uncertainty range in the metallicity and belongs to spaxels with lower S/N.

{The total stellar mass of this merging system is log(M$_{\star}$/M$_{\odot}$) $\sim$ 9.7 (see Table~\ref{Pair_properties}). In a similar mass range, \cite{2020Curti} estimated a value of 12 + log(O/H) = 8.6 dex for star-forming dwarf galaxies using SDSS data. This estimate, similar to ours, is based on the N${\rm 2}$ empirical calibrator from \cite{2004Pettini} and is approximately 0.05 dex higher than the measurements we obtained from the integrated spectrum of this merging pair, as well as each component individually. When employing the O3N2 index from \cite{2004Pettini}, we find an average 12 + log(O/H) = 8.40 dex for this component, which is still 0.2 dex below \cite{2020Curti}'s findings for star-forming dwarfs in comparable mass ranges using similar calibration methods. }

However, these discrepancies do not necessarily imply that the merger event leads to {lower} gas-phase metallicity in dwarf galaxies, mainly since the 0.05 dex difference we report falls within the typical systematic uncertainty of approximately 0.2 dex associated with the empirical calibrations employed \citep{2004Pettini}. Moreover, further direct comparisons with other interacting or merging dwarf systems are not possible since analysis as such is also limited in the literature and existing studies have employed varied methods for deriving gas-phase metallicity \citep[e.g.][]{2008Michel-Dansac}.

\subsection{Star formation}\label{Results5}

\begin{figure}
\centering
\includegraphics[scale=0.8]{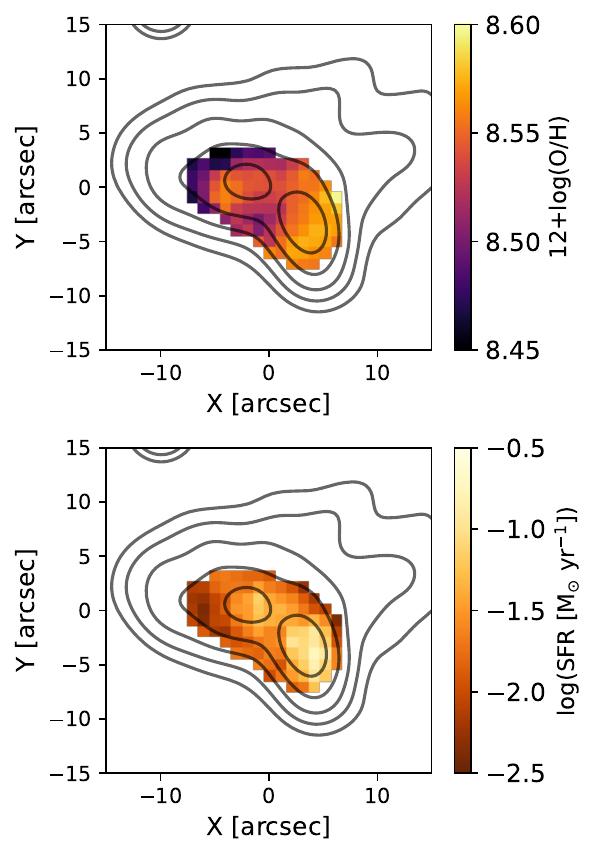}
\caption{Gas-metallicity and SFR maps. \textit{Top panel:} Map of oxygen abundance as a proxy of gas-phase metallicity. \textit{Bottom panel:} Map of SFR. Only spaxels with S/N$>$15, {similar to Fig.~\ref{CAVITY35844_gas_maps}}, are shown here. Contours are the same as in Fig.~\ref{Fig2}.}
\label{CAVITY35844_dust_metal_SFR}
\end{figure}

Using the extinction corrected H$\alpha$ flux, we measured the SFR in this merging system {following the equation of \cite{1998Kennicutt}:
\begin{equation}
\rm SFR(M_{\odot} yr^{-1}) = 7.9\times 10^{-42} L(H\alpha), 
\end{equation}}
where L(H$\alpha$) is the luminosity of the H$\alpha$ emission line (in erg/s), measured at the comoving radial distance of 124.3 Mpc \citep{2012Pan}. On the lower panel in Fig.~\ref{CAVITY35844_dust_metal_SFR} we show log(SFR) values for each spaxel. The average log(SFR [M$_\odot \rm yr^{-1}$]) over all the spaxels with S/N $>$ 15 in this system is -1.47. The integrated SFR, derived from the integrated spectra of the entire system, is log(SFR~[M$_\odot \rm yr^{-1}$])$=$0.61. {Based on the unresolved 1.4 GHz radio-continuum flux measurements reported for this system \citep{2015Helfand,2018deGasperin}, we derived a log(SFR [M$_\odot \rm yr^{-1}$]) = 1.05 using the calibration provided by \cite{2009Kennicutt}. As expected, the log(SFR) derived from the radio-continuum is 0.4 dex higher than that obtained from the H$\alpha$ emission line, which is likely due to the reduced impact of dust extinction in the radio regime. }The integrated log(SFR [M$_\odot \rm yr^{-1}$]) measured for components A and B from their integrated spectra are 0.38 and 0.15, respectively. Both integrated SFR and the spatial map presented in the lower panel of Fig.~\ref{CAVITY35844_dust_metal_SFR} indicate that component A has a higher SFR compared to component B (see also Table~\ref{Gas_properties}). 

The integrated SFR of this interacting system, whether considered as a whole or for each merging component individually, exceeds the estimates provided by \citet{2021Vilella-Rojo} for star-forming dwarf galaxies with comparable stellar masses, where log(SFR [M$_{\odot}$ yr$^{-1}$]), as they reported, is $\sim$ $-0.5$, derived using H$\alpha$ emission lines from J-PLUS data \citep[a similar value was also reported using SDSS fibre spectra by][]{2017DuartePuertas}. Both cited studies rely on one-dimensional spectroscopy and apply methodologies similar to ours in deriving the SFR, which aids in minimising potential systematic differences when compared with the integrated SFR measured here. These findings are in line with other studies showing that the interaction between two dwarf galaxies can increase their SFR \citep[e.g.][]{2015Stierwalt,2023Gao,2024Subramanian}.

{Using the stellar mass estimates, we derived the specific SFR (sSFR = SFR/$M_{\star}$) of the total system, as well as each merging component (see Table~\ref{integrated_values}). The definition of the star formation main sequence in \cite{2020Janowiecki}, places this system in the starburst category, a typical occurrence in gas-rich (wet) mergers \citep[e.g.][]{1995Hibbard}. These findings are also consistent with \cite{2012Lambas} who showed that at a given stellar mass, major mergers are more efficient in forming new stars. }

As with the gas-phase metallicity, a direct comparison with the SFR of other merging dwarf systems was not feasible. This is because SFR and sSFR values for many merging and interacting dwarf galaxies are either not reported in the literature or are derived using different methods or indicators (e.g. H$\alpha$ vs. UV), rendering meaningful comparisons with our case impractical.

\section{Discussion}\label{Discussion}

In this work, we analysed the physical properties of CAVITY35843 and CAVITY35844, a binary merger event of dwarf galaxies. This system is among the very few, if not the first, merging dwarf pairs located in the central region of a cosmic void investigated to date. The unique configuration of this pair allowed for a detailed analysis of its gas properties and kinematics. In what follows, we discuss our results and compare this system with other interacting dwarfs studied in the literature. 

\subsection{Main properties of the interacting pair}
Our analysis of the gas component, traced by the H$\alpha$ emission line in both components, reveals clear signs of rotation in each merging galaxy. This rotation is also evident in their stellar components, suggesting that both merging dwarfs likely possessed rotating discs prior to the merger event. Based on the presence of rotation in each component, along with their high SFR, gas-phase metallicity, and ionised gas, we speculate that these merging pairs were initially star-forming dwarf galaxies with rotating discs. Dwarf galaxies with these characteristics are commonly found in voids \citep[e.g.][]{2011Kreckel,2023A&A...680A.111D}. 

Analysis of the rotation velocity profiles in Section~\ref{Result2} indicates that both galaxies have comparable ranges of dynamical mass up to {2.9 kpc from their centres} (Fig. \ref{Fig4}). Additionally, the nearly identical gas-phase metallicities derived from the integrated spectra of each component, when considered alongside the well-established mass-metallicity relations \citep[e.g.][]{2020Curti}, suggest that the two merging galaxies likely had similar masses. {This conclusion is further supported by our stellar mass estimates for each component, based on the fits to the IFU data. Taken together, these findings point to a nearly 1:1 merger event in progress.}

{ The total dynamical mass of this merging pair, based on analysis presented in Fig.~\ref{Fig4} is log(M$_{\rm dyn}$/M$_{\odot}$) = 10.22. \cite{2020Durbala} reported a total flux of 1.8 [Jy km~s$^{-1}$] for the HI neutral gas in this system, corresponding to log(M$_{\rm HI}$/M$_{\odot}$) = 9.86 $\pm$ 0.05, that is 2.2 times smaller than the dynamical mass derived here. We estimated a total stellar mass of log(M$_{\star}$/M$_{\odot}$) $\sim$ 9.7. This value is approximately 1.5 times smaller than the HI mass, making this system possibly one of the rare cases with high abundance of neutral gas studied to date \citep[e.g.][]{2018Paudel}.  It is essential to note that uncertainties in the stellar mass estimates and dynamical mass measurements, as discussed in Section~\ref{Result3}, may impact this comparison. Additionally, the HI profile of this interacting pair, as analysed by \cite{2020Durbala}, is incomplete due to missing channels, introducing further uncertainty in the estimated HI mass.}

{In Section~\ref{Results1}, we discussed that the gas and stellar components in this merging system are co-rotating in their structures. The individual maps in Fig.~\ref{Fig4} suggest that the two components are possibly counter-rotating relative to each other (i.e. component B is retrograde with respect to component A), but confirming this would require knowledge of their 3D orientation on the sky. In retrograde encounters such as the one studied here, tidal effects persist for a shorter duration, leading to less extended tidal features. Conversely, major mergers (i.e. mass ratios smaller than 3:1) tend to produce more extended tidal structures that endure longer \citep[e.g.][]{1972Toomre,2013Duc}. Given that this system is a retrograde merger with a mass ratio of $\sim$ 1:1, it remains unclear which of these factors has played a dominant role in shaping the observed tidal tails in Fig.~\ref{Fig1}. While resolving this question is beyond the capabilities of our data and the scope of this work, this system serves as an excellent textbook case of a dwarf-dwarf merger, providing an opportunity for modellers to explore the parameter space and assess the impact of internal and orbital dynamics on the formation and evolution of tidal features.}

Further analysis of the gas component, whether for the system as a whole or for each individual merging component, reveals an elevated SFR compared to other star-forming dwarf galaxies with similar stellar masses. Specifically, the SFR measured for component A is nearly twice that of component B. Both systems are currently undergoing a starburst phase, commonly associated with mergers between gas-rich dwarf galaxies possessing extended HI discs \citep{2008Bekki}. During such events, the time-dependent tidal gravitational field efficiently funnels gas into the central regions of the merging galaxies, where the resulting increase in gas density triggers intense star formation. The \textit{g-r} colour from INT imaging data (shown in Panel E of Fig. \ref{Fig1}) reveals additional possible starburst regions in the eastern part of component B and along the system’s tidal tail. These starburst regions also are likely the result of local gas over-densities caused by the interaction.

The gas-phase metallicity derived for the entire system and each merging component is {lower than} what is typically reported for dwarf galaxies of similar stellar masses. This is {expected}, as recent studies have often found lower gas-phase metallicities, especially in the central regions of galaxies undergoing major mergers or interactions \citep[e.g.][]{2019Thorp, 2024Garay-Solis}. On the other hand, there are a few studies that report higher gas-phase metallicity, by 0.2 dex, in merging and interacting galaxies, particularly in low-mass ones \citep[e.g.][]{2008Michel-Dansac,2022Porter}. However, the difference between the metallicity measured in this study and values reported for star-forming (non-interacting) dwarf galaxies falls within the systematic uncertainties of approximately 0.2 dex associated with empirical calibrations. As such, no definitive conclusions can be drawn. A systematic analysis of a larger sample of interacting or merging dwarf galaxies, particularly in void environments, is necessary to validate these findings.

We also observed a distinctive distribution of red colours, appearing as two small, connected arcs separating the main body of the galaxies from the faint tidal tail. This red colour distribution (Panel D of Fig. \ref{Fig1}), when compared with the dust distribution visible in the DECaLS images (Fig. \ref{Fig1}), suggests that it is primarily associated with dust in this interacting system, that is also influenced by the ongoing merger. {The dust observed can be related to the ionised gas or the molecular or atomic gas compressed by the merger activity.} However, we could not detect this feature in the extinction maps derived from the IFU data, likely due to its lower spatial resolution compared to the INT imaging.

\begin{figure}
\centering
\includegraphics[scale=0.55]{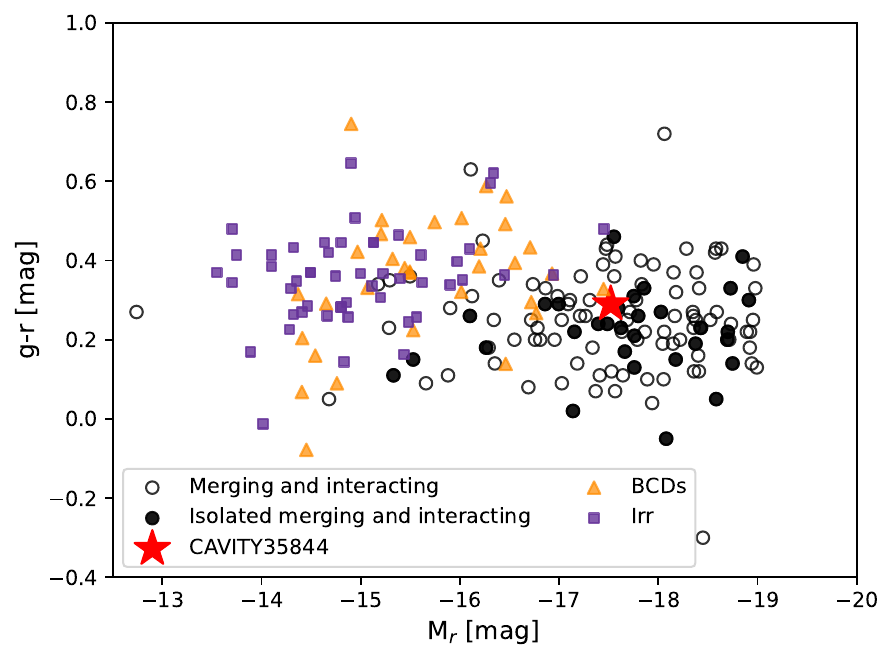}
\includegraphics[scale= 0.55]{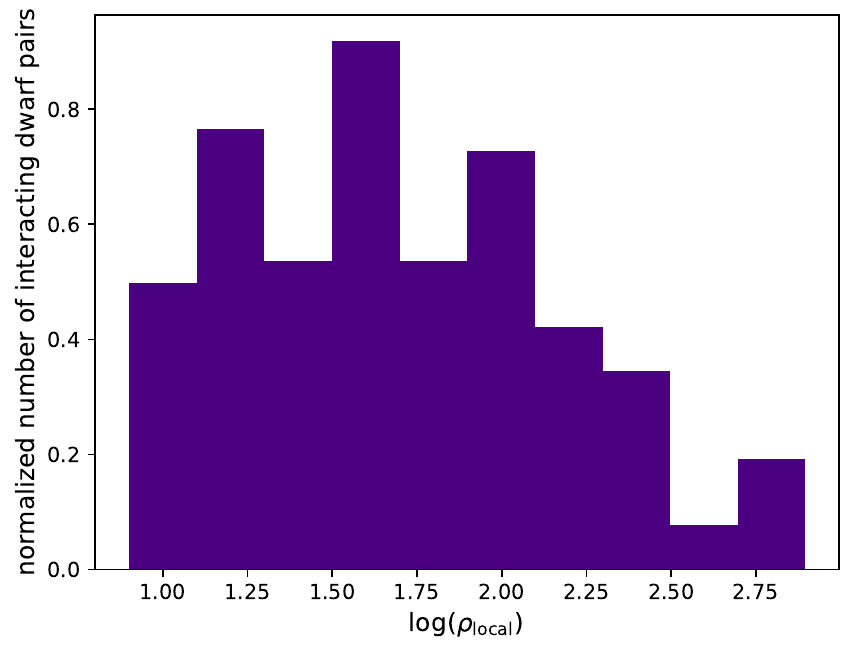}
\caption{Comparison with other merging systems. \textit{Top panel:} Comparison of the \textit{g-r} colour of interacting or merging dwarf galaxies from the \protect\cite{2018Paudel} sample (shown with filled and empty black circle symbols for isolated and non-isolated cases, respectively) with the interacting system (shown as red data point), the \protect\cite{2006Lisker} sample of BCDs (orange triangles), and the \protect\cite{2014Pak} sample of irregular dwarf galaxies (shown in purple squares). 
\textit{Bottom panel:}  Distribution of normalised projected number density shown for the galaxies in the \protect\cite{2018Paudel} sample with more than three neighbours. The local density was computed as the number of neighbours found within a projected distance of 1.5 Mpc and a relative radial velocity of $\Delta V$ = 500 km~s$^{-1}$ divided by the volume. }
\label{discussion}
\end{figure}

In Fig. \ref{discussion}, we compare the \textit{g-r} colour of this interacting system with other interacting/merging dwarf galaxies, represented by circle symbols. The comparison is made with the sample from \cite{2018Paudel}, which consists of 177 nearby merging dwarf galaxies with log(M$_{\star}$/M$_{\odot}$) $<$ 10. We selected this sample as our reference because it is the largest catalogue of merging dwarf galaxies available in the literature to date, being more than 1.5 times larger than the TiNy Titans catalogue \citep{2015Stierwalt}. We present the average \textit{g-r} value of the merging system discussed here, calculated using INT-convolved data (similar to Panel A of Fig. \ref{Fig2}), and find no significant difference between it and the other merging or interacting dwarf galaxies from \cite{2018Paudel} in terms of \textit{g-r} colour. In fact, all these interacting systems exhibit similar colours to those of BCDs (shown in blue; \cite{2006Lisker}) and dwarf irregular galaxies (shown in purple; \cite{2014Pak}), both of which are subclasses of dwarf galaxies characterised by high SFR.

Some studies predict that after coalescence and in the absence of strong environmental influences, merging dwarf galaxies can evolve into more massive, star-forming dwarfs with irregular shapes or even become BCDs \citep[e.g.][]{2008Bekki, 2020Kado-Fong, 2020Zhang}.  {Other research suggests that, even after a major merger of progenitor galaxies with gas-rich discs, as seen in CAVITY35843 and CAVITY35844, disc-like morphologies can still emerge in the merger remnant. This outcome is possible only if tidal torquing is less dominant and the gas retains its high angular momentum, allowing it to settle into a disc at the end of the merger \citep[e.g.][]{2005Springel,2009Hopkins}. Further investigations into these scenarios for the merging system studied here would require detailed modelling of the merging event, which is beyond the scope of this work.}

\subsection{Environment of the interacting system}

CAVITY35843 and CAVITY35844 form a merging system of two dwarf galaxies located in the centre of a cosmic void. These two dwarfs are located at the distance of 0.13 R$_{\rm e,void}$ from their host void centre, where the R$_{\rm e,void}$ is the effective radius of the void, assuming it is a sphere with the same volume as the actual void. Our analysis of their local environment using the velocity difference-projected distance space method \citep{2015A&A...578A.110A} reveals that this merging system has no massive neighbours. Specifically, we examined their local environment using the NSA-Sloan catalogue (NSA\footnote{\url{https://www.sdss4.org/dr17/manga/manga-target-selection/nsa/}}), within a projected distance of 1.5 Mpc and a relative radial velocity of $\Delta V$~=~500 km~s$^{-1}$. The only galaxy found within this range is CAVITY38680, a star-forming dwarf galaxy with a stellar mass of log(M$_{\star}$/M$_{\odot}$)~=~8.88, located at a projected distance of 0.343 Mpc from the merging system (see Fig. \ref{Fig1}).

To assess potential environmental effects on the properties of merging and interacting dwarf galaxies, we quantified the local environment of galaxies in the \cite{2018Paudel} sample. This analysis is necessary due to the limited information available regarding the local or large-scale environments of these systems. We adopted similar criteria as in \cite{2015A&A...578A.110A}, expanding the search area to a projected distance of 1.5 Mpc and considering a relative radial velocity of $\Delta V$ = 500 km~s$^{-1}$. We found that only 17 (out of 177)  galaxies in their sample are in complete isolation, with no neighbouring galaxy brighter than M$_{\rm r}$ $\sim$ -17 mag within the searched space. Additionally, we identified 29 dwarf interacting pairs in the sample that have only one neighbouring galaxy, within the investigated velocity difference-projected distance space. These galaxies, along with the isolated ones, are represented by filled circle symbols in the top panel of Fig. \ref{discussion}. We could not detect any trend between the local projected density and the \textit{g-r} colour of these galaxies. 

We excluded these galaxies from the sample and computed the projected number density of the neighbouring galaxies of each interacting dwarf within a distance of 1.5 Mpc {($\rho_{\rm local}$)}. The distribution of these local number densities is shown in the lower panel of Fig \ref{discussion}. We found that the majority of this sample is located in locally populated regions, with more than three neighbouring galaxies, resembling `group-like' environments. This aligns with previous studies concluding that galaxy mergers are more common in a group-like environment, where galaxies are not as isolated as in voids but also are not moving relatively as fast as those in clusters \citep[e.g.][]{2008Tran, 2012Alonso, 2014Deason, 2018Paudel, 2024Sureshkumar}. 

In their study, \cite{2018Paudel} concluded that dwarf-dwarf interactions and mergers occur frequently in low-density environments, based on a search within a 700 kpc projected distance and 700 km~s$^{-1}$ relative velocity difference. While this might appear to contradict the present findings, it actually highlights a bias due to different definitions of local environments for galaxies. As \cite{2023Omori} discussed, the dependence of merger rates on the galaxy environment varies with the scale at which the environment is defined. Our study considered a more extensive velocity difference-projected distance space and lower-mass galaxies as neighbours. Hence, our results indicate that although most galaxies in the \cite{2018Paudel} sample do not have massive neighbours (log(M$_{\star}$/M$_{\odot}$) > 10) within a projected distance of 700 kpc, they are not completely isolated. 

Based on our definition of the local environment and isolation, only 1\% of the \cite{2018Paudel} sample has conditions similar to this merging system, meaning they are nearly 1:1 merging events. Interestingly, most of the isolated merging and interacting cases listed in that study for which the relative mass ratio of merging components (m$_{1}$/m$_{2}$) has been reported (10 out of 18 isolated cases) fall in the range of m$_{1}$/m$_{2}$ $\geq$ 5 making them mostly minor merger events \citep{2008Michel-Dansac,2018Paudel}. In these cases, the less massive galaxies, usually the dwarfs, merge with the more massive companion, likely due to the larger gravitational forces exerted by the latter. This is not the case for CAVITY35843 and 35844.

The trigger for the merging event between the two dwarf galaxies in this system remains an open question. Most merging events, as discussed before, occur in group-like environments, where high galaxy number density and the tidal field of the host halo promote such interactions. Additionally, isolated mergers typically involve two galaxies with a significant mass difference, where the more massive galaxy traps the less massive one in its gravitational field. However, components A and B in this interacting pair do not follow these patterns. The merging we witness can be due to the global dynamics of the void and the distribution of dark matter within it \citep[e.g.][]{2013Beygu,2013Rieder,2023Courtois} which shape the flow of void galaxies. {\cite{2017Chengalur} proposed that the wet mergers in low-density environments, such as voids, take place as a consequence of structure formation, which is more slowly at small scales than in regions with average density. This can also be a possible pathway for the formation of gas-rich discs.} Similarly, the past three-body encounters between components of merging system and their nearby dwarf galaxy might be responsible for triggering this merger event. {However, the INT deep imaging utilised in this study reveals no tidal streams or features indicative of past three-body encounters. }  Further studies on the dynamical evolution of voids and mergers involving three-body interactions are needed to evaluate these scenarios.

\section{Summary}\label{Summary}

In this study, we examined the kinematics and gas properties of CAVITY35843 and CAVITY35844, a dwarf-dwarf merging pair near the centre of a cosmic void. This system is remarkable due to its orientation relative to the line of sight and its unique local and large-scale environment. As an interacting system, it presents a compelling opportunity to explore the interactions typical of gas-rich low-mass galaxies at high redshifts, that are fundamental to the hierarchical galaxy formation model. The main results of this study are summarised below:
\begin{itemize}
\item CAVITY35843 and CAVITY35844 form a nearly 1:1 dwarf-dwarf merging system located near the centre of a cosmic void (R = 0.13 R$_{\rm e,void}$), one of the least dense region of the local Universe. Identifying such a system in a void is atypical, as most dwarf-dwarf merging systems are found in group-like environments, where the balance between galaxy density and relative velocities facilitates merging events.

\item We speculate that past three-body interactions with the only neighbouring dwarf galaxy identified for this system, CAVITY38680, might have triggered the merger. However, the deep INT imaging detected no tidal streams or peculiar features indicative of such interactions. Alternatively, the dynamical evolution of voids and their associated dark matter structures, which govern the flow of galaxies, could also be responsible for initiating this merging event.

\item Both components of this merging system are undergoing starburst activity but exhibit slightly lower gas-phase metallicities compared to non-interacting star-forming dwarf galaxies of similar stellar masses, though these values remain within the systematic uncertainties of the measurement indicators. 

\item {While some theoretical studies suggest that major mergers can trigger AGN activity \citep[e.g.][]{2017Capelo}, our analysis using the BPT diagram and the \cite{2011CidFernandes} detection method based on EW(H$\alpha$) finds no evidence of AGN activity in either component of this merging pair. However, whether this absence of AGN activity is due to the low-mass nature of the progenitors or the specific merger stage remains uncertain. Some studies suggest that AGN activity peaks in post-coalescence mergers \citep[e.g.][]{2014Satyapal}, but such trends have only been observed in massive merging galaxies \citep[i.e. with log(M$_{\star}$/M$_{\odot}$) > 11.0;][]{2024Comerford}. A more extensive analysis of merging dwarf galaxies at different evolutionary stages is needed to better understand the role of mergers in AGN triggering.}

\item {We estimate a neutral gas-to-stellar mass ratio of approximately 1.5, positioning this system among the rare cases exhibiting high neutral gas abundances observed to date \citep[e.g.][]{2018Paudel, 2018Ellison}. While previous studies indicate that dwarf galaxies in low-density environments, such as voids, are generally gas-dominated, our findings further suggest that the neutral gas in this system has persisted through the merging event (at least up to the current stage). High-resolution follow-up investigations of the molecular gas content in this and similar void galaxies are crucial for gaining deeper insight into the complex processes governing mergers.
}

\item We found no significant differences between the \textit{g-r} colour of this merging system, located at the centre of a void, and other merging dwarf systems situated in different environments, including group settings. This suggests that photometric colour is primarily driven by the merging event, with environmental conditions playing a secondary role during this phase. However, we could not evaluate this trend for other physical properties. Direct comparison of the gas-phase metallicity and SFR of this system with similar cases in the literature was not possible due to the limited number of references available and the differing data characteristics and methodologies, which precluded meaningful comparisons.

\end{itemize}

These results are based on the detailed analysis of a single 1:1 merging dwarf pair located in a void. While this study provides valuable insights, it represents just one case among many that remain to be explored in order to fully grasp the broader properties of dwarf-dwarf mergers across diverse environments. To validate and expand on these findings, future research should target statistically significant samples with well-characterised environmental contexts, utilising deep IFU spectroscopy and high-quality photometry for more comprehensive analyses.

\begin{acknowledgements}
We thank the referee for their thoughtful comments, which have improved the content and presentation of this paper. 
This paper is based on data obtained by the CAVITY project, funded by the Spanish Ministry of
Science and Innovation under grants PID2020-113689GB-I00 and PID2023-149578NB-I00 as well as by Consejería
de Universidad, Investigación e Innovación and Gobierno de España and
Unión Europea - NextGenerationEU through grant AST22\_4.4, financed by MCIN/AEI/10.13039/501100011033, the project A-FQM-510-UGR20 financed
from FEDER/Junta de Andalucía-Consejería de Transformación Económica, Industria, Conocimiento y Universidades/Proyecto and by the grants P20-00334 and FQM108, financed by the Junta de Andalucía (Spain). Based on observations collected at Centro Astronómico Hispano en Andalucía (CAHA) at Calar Alto, proposal F21-3.5-101, operated jointly by Junta de Andalucía and Consejo Superior de Investigaciones Científicas (IAA-CSIC).
BB acknowledges financial support from the Grant AST22\_4.4, funded by Consejería de Universidad, Investigación e Innovación and Gobierno de España and Unión Europea – NextGenerationEU, and by the research project PID2020-113689GB-I00 and PID2023-149578NB-I00 financed by MCIN/AEI/10.13039/501100011033.
SBD acknowledges financial support from the grant AST22.4.4, funded by Consejer\'ia de Universidad, Investigaci\'on e Innovaci\'on and Gobierno de Espa\~na and Uni\'on Europea -- NextGenerationEU, also funded by PID2020-113689GB-I00, financed by MCIN/AEI, and the support of the Spanish Ministry of Science, Innovation and Universities through the project PID--2021--122544NB--C43.
IP  acknowledges financial support from the grant AST22.4.4, funded by Consejería de Universidad, Investigación e Innovación and Gobierno de España and Unión Europea — NextGenerationEU, and by PID2023-149578NB-I00, financed by MCIN/AEI.  
AZ, UL, EF, LSM, MAF and MR acknowledge support from projects PID2023-150178NB-I00 and PID2020-114414GB-100, financed by MCIN/AEI/10.13039/501100011033. 
J.R. acknowledges financial support from the Spanish Ministry of Science and Innovation through the project PID2022-138896NB-C55.
M.A-F. acknowledges support from the Emergia program (EMERGIA20\_38888) from Consejer\'ia de Universidad, Investigaci\'on e Innovaci\'on de la Junta de Andaluc\'ia.
DE acknowledges support from a Beatriz Galindo senior fellowship (BG20/00224) from the Spanish Ministry of Science and Innovation.
TRL acknowledges support from Juan de la Cierva fellowship (IJC2020-043742-I) and Ram\'on y Cajal fellowship (RYC2023-043063-I, financed by MCIU/AEI/10.13039/501100011033 and by the FSE+).
LSM acknowledges support from Juan de la Cierva fellowship (IJC2019- 041527-I).
RGB, acknowledges financial support from the Severo Ochoa grant CEX2021-001131-S funded by MCIN/AEI/ 10.13039/501100011033 and to grant PID2022-141755NB-I00.
AFM has received support from RYC2021-031099-I and PID2021-123313NA-I00 of MICIN/AEI/10.13039/501100011033/FEDER,UE, NextGenerationEU/PRT.
SDP acknowledges financial support from Juan de la Cierva Formaci\'on fellowship (FJC2021-047523-I) financed by MCIN/AEI/10.13039/501100011033 and by the European Union `NextGenerationEU'/PRTR, Ministerio de Econom\'ia y Competitividad under grants PID2019-107408GB-C44, PID2022-136598NB-C32, and is grateful to the Natural Sciences and Engineering Research Council of Canada, the Fonds de Recherche du Qu\'ebec, and the Canada Foundation for Innovation for funding. 
JFB acknowledges support from the PID2022-140869NB-I00 grant from the Spanish Ministry of Science and Innovation. 
EF acknowledge support from the Junta de Andaluc\'ia (Spain) local government through the FQM-108 project.
GTR acknowledges financial support from the research project PRE2021-098736, funded by MCIN/AEI/10.13039/501100011033 and FSE+.
This research made use of astropy, a community-developed core python \citep[http://www.python.org,][]{python} package for Astronomy \citep{2013A&A...558A..33A,2018AJ....156..123A,2022ApJ...935..167A}; IPython \citep{PER-GRA:2007}; Matplotlib \citep{Hunter:2007}; NumPy \citep{2011CSE....13b..22V}; SciPy \citep{2020SciPy-NMeth,scipy_11255513} and Pandas \citep{mckinneyprocscipy2010}. This research has made use of the NASA/IPAC Extragalactic Database, operated by the Jet Propulsion Laboratory
of the California Institute of Technology, under contract with the National Aeronautics and Space Administration. Funding for SDSS-III has been provided by the Alfred P. Sloan Foundation, the Participating Institutions, the National Science Foundation, and the U.S. Department of Energy Office of Science. The SDSS-III web site is \href{http://www.sdss3.org/}{http://www.sdss3.org/}. The SDSS-IV site is \href{http://www.sdss.org}{http://www.sdss.org}.

\end{acknowledgements}

\bibliographystyle{aa} 
\bibliography{main}

\begin{appendix}
\onecolumn
\section{More on the kinematics of the system}\label{AppA}
\begin{figure*}
\centering
\includegraphics[width=\textwidth]{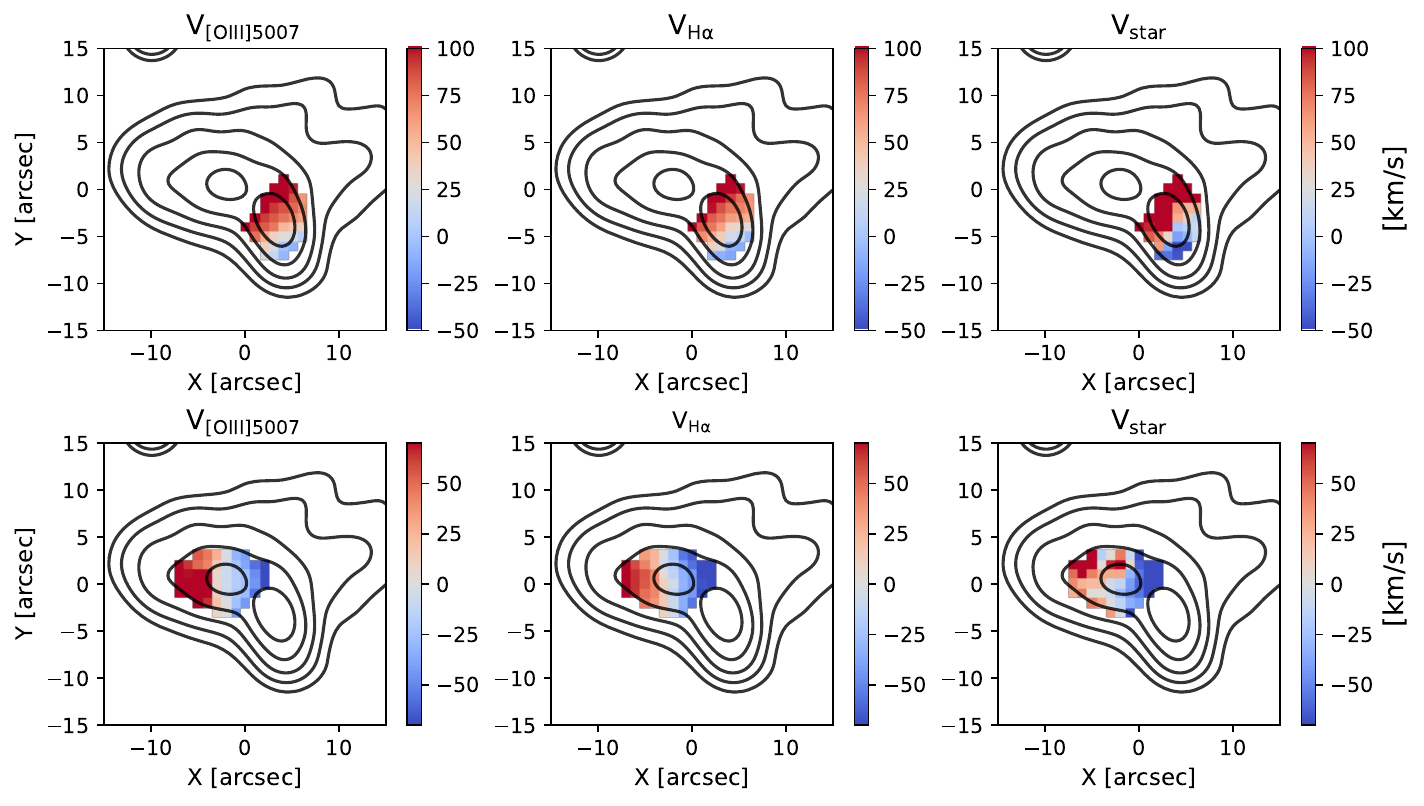}
\caption{Gas and stellar velocity maps of merging component A (top row) and component B (bottom row), taking into account the $\sim$ 100 km~s$^{-1}$ difference in the relative velocity. }
\label{CAVITY35844_Kinematics_AppendixA}
\end{figure*}

As discussed in Section~\ref{Results1}, a lopsided rotation is observed in both the gas and stellar radial velocity maps of this merging system. Part of this lopsidedness can be attributed to the $\sim$100~km~s$^{-1}$ difference in the relative velocity between the merging components. Additionally, the higher $\sigma_{\rm gas}$ values in a small linear region between the components, as previously mentioned, are likely due to the emission lines from both components overlapping in a single spaxel, that are not resolved by the PPAK IFU's spectral and spatial resolution. However, this region serves as a useful separation line between the components. In Fig.~\ref{CAVITY35844_Kinematics_AppendixA}, we show the radial velocity maps of each merging component, separated by the elevated $\sigma_{\rm gas}$ line. After accounting for the radial velocity difference, component B exhibits clear rotation. Component A also shows signs of rotation, although less clearly, possibly due to its alignment along the line of sight and the significant effects of the interaction on its kinematics. 

\begin{figure*}
\centering
\includegraphics[width=\textwidth]{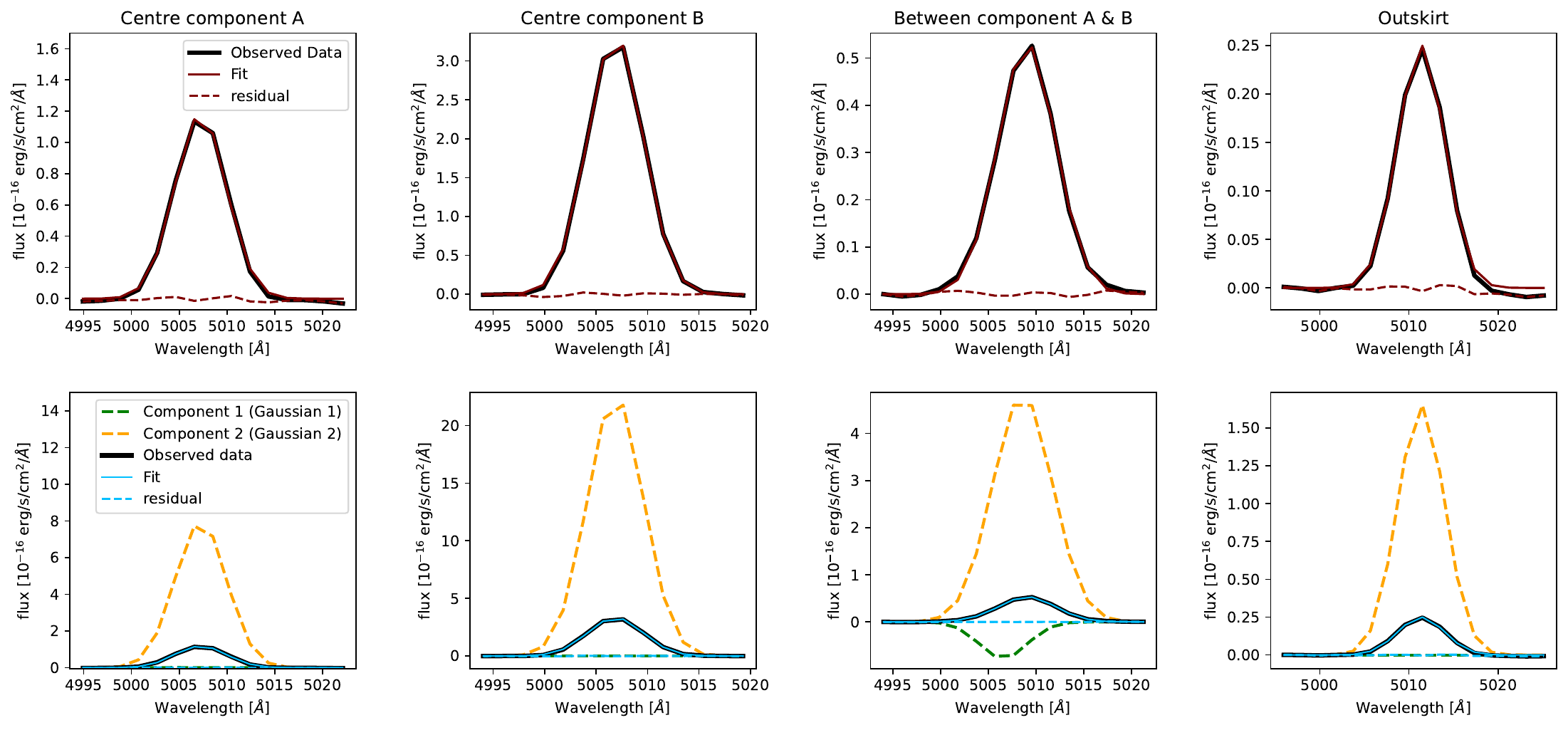}
\caption{Examples of fits for the [OIII]$\lambda$5007 emission line shown for four distinct regions, arranged from left to right: component A, component B, the region between the two components, and the outskirts of the merging pair. The top row illustrates single Gaussian fits, while the bottom row showcases examples of double Gaussian fits.  }
\label{CAVITY35844_lines_AppendixA}
\end{figure*}

In Fig.~\ref{CAVITY35844_lines_AppendixA}, we display the [OIII]$\lambda$5007 emission line for four spaxels from different spatial regions of the interacting system, namely (from left to right): the centre of component A, the centre of component B, the region between the interacting components, and the outskirts of component B, where we observe elevated $\sigma_{\rm gas}$ in the kinematic maps of Fig.\ref{CAVITY35844_Kinematics}. The top row shows the results of a single Gaussian fit to the emission line, along with the residuals of the fits. The bottom row presents the results of a double Gaussian fit to the emission line, which failed in all four cases. We were unable to detect any visible contributions from double Gaussian components in the gas emission lines detected with the PPAK-IFU data. This limitation is primarily due to the instrument's resolution, which is insufficiently sensitive to resolve such cases.

\section{${\tt ^{3D}BAROLO}$ parameters}\label{AppB}
In Table~\ref{BAROLO_parameters_table}, we present the input {and output} parameters used for the ${\tt ^{3D}BAROLO}$ fitting for each merging component, as described in Section \ref{Results3}. The details regarding the parameters and their meaning can be found in the documentation page of ${\tt ^{3D}BAROLO}$\footnote{\url{https://bbarolo.readthedocs.io/en/latest/tasks/fit3d.html\#parameters}}. {Here, we assumed that the position of the centre(\texttt{xpos}, \texttt{ypos}), the scale-height of the disc (\texttt{z0}), and the systemic velocity (\texttt{vsys}) remain constant across the discs of the galaxies. To simplify the fitting process, we provided initial values for these parameters to ${\tt ^{3D}BAROLO}$, and kept them fixed during the fitting. For \texttt{vsys}, we did not specify a value; instead, it was estimated by the software as the central velocity from the global line profile.}

\setlength{\tabcolsep}{8.pt}
\label{BAROLO_parameters}
\begin{table*}
\caption{\label{BAROLO_parameters_table} ${\tt ^{3D}BAROLO}$ input parameters.}
\centering
\begin{tabular}{c c c c }
\hline
Parameter & Value - component A & Value - component B & Comments \\
\hline
\hline
\texttt{nradii}             & 25                & 16                & Fine-tuning \\ 
\texttt{radsep}             & 0.8               & 1.0               & Datacube limitations \\ 
\texttt{xpos*}              & 16.2              & 11.0              & {Centre [pix]}, isophote fitting \\ 
\texttt{ypos*}              & 7.5               & 12.0              & {Centre [pix]}, isophote fitting \\ 
\texttt{vsys*}              & -                 & -                 & {Systematic velocity [$ {\rm km\ s^{-1}}$]}, estimated by the software \\ 
\texttt{z0*}                & 1                 & 1                 & {The scale-height of the disk [arcsec]}, default value \\ 
\texttt{vrot}               & 75                & 150               & {Rotational velocity [$ {\rm  km\ s^{-1}}$]}, section~\ref{Results1} \\ 
\texttt{vdisp}              & 80                & 80                & {Velocity dispersion [$ {\rm km\ s^{-1}}$]}, section~\ref{Results1} \\ 
\texttt{inc}                & 43                & 50                & {Inclination [deg]}, isophote fitting \\ 
\texttt{pa}                 & 11                & 60                & {Position angle [deg]}, isophote fitting\\ 
\texttt{mask}               & THRESHOLD=0.2     & SEARCH            & Datacube limitations \\ 
\texttt{growthcut}          & 2.5               & 2.5               & Default value \\ 
\texttt{free}               & VROT VDISP PA INC & VROT VDISP PA INC &   \\ 
\texttt{norm}               & AZIM              & AZIM              & Default or recommended value \\ 
\texttt{restwave}           & 6562.8            & 6562.8            & H$\alpha$ rest-frame wavelength \\ 
\texttt{redshift}           & 0                 & 0                 & Default or recommended value \\ 
\texttt{linear}             & -                 & -                 & Default or recommended value \\ 
\texttt{bweight}            & 0                 & 0                 & Default or recommended value \\ 
\texttt{normalcube}         & FALSE             & FALSE             & Default or recommended value \\ 
\hline 
\texttt{vrot}               & 81.8              & 164.2             &  \\ 
\texttt{vdisp}                 & 78.4              & 78.8              &  \\ 
\texttt{inc}              & 43                & 50                &  \\ 
\texttt{pa}         & 43.8              & 82.5              &  \\ 
\hline
\end{tabular}

\vspace{0.2cm} 
\noindent
Input and output parameters of ${\tt ^{3D}BAROLO}$. The columns display the names of both the input {and output} parameters, along with the corresponding values for components A and B, and relevant comments for each parameter. {Parameters that were fixed during the fit are marked with an asterisk and are listed in the first part of the table, alongside other input parameters. The second part of the table presents the geometrical parameters derived from the fit results.}
The comments are categorised into four general groups: Fine-tuning are parameters estimated after several iterations. Datacube limitations are parameters determined as the best value for the datacube. Isophote fitting are parameters estimated using isophote fitting. Default or recommended value are parameters set as default or recommended by the pipeline.

\end{table*}

\section{Source of ionisation}\label{AppC}
\begin{figure}
\centering
\includegraphics[scale = 0.8]{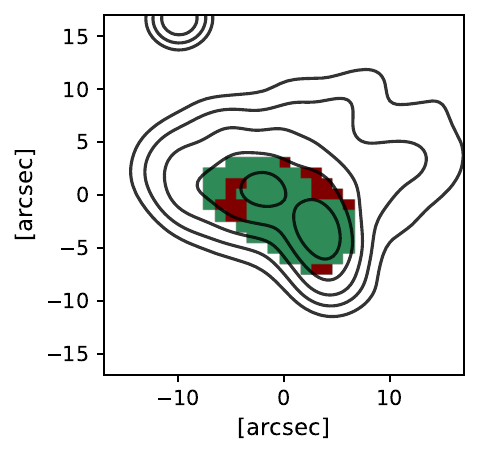}
\caption{Spatial distribution of spaxels within the system where the spaxels in the AGN-dominated ionisation zone of Fig.~\ref{CAVITY35844_BPT} are marked in red, and those in the star-forming region are marked in green }
\label{BPT_appendix}
\end{figure}
In Section~\ref{Result3} and Fig.~\ref{CAVITY35844_BPT}, we demonstrated that some spaxels in this merging system fall within the region of the BPT diagram typically associated with AGN-dominated ionisation. Fig.~\ref{BPT_appendix} shows the spatial distribution of these spaxels within the system, where the spaxels in the AGN-dominated ionisation zone are marked in red, and those in the star-forming region are marked in green. Upon comparison with the S/N map and the results presented in Figs.~\ref{CAVITY35844_Kinematics} and \ref{CAVITY35844_gas_maps}, it is clear that the spaxels in the AGN zone exhibit low S/N and are situated in the outer regions of both merging components, where detecting AGN signatures or black holes is not feasible.

\end{appendix}

\end{document}